\documentclass[reprint, amsmath,amssymb, aps,pra]{revtex4-1}

\usepackage{amsmath}
\usepackage{amsthm}
\usepackage{amssymb}
\usepackage[breaklinks,colorlinks]{hyperref}
\usepackage{bm}
\usepackage{dcolumn}
\usepackage{enumerate}
\usepackage{enumitem}
\usepackage{graphicx}
\usepackage{latexsym}
\usepackage{mathtools}
\usepackage{url}
\usepackage{fullpage}
\usepackage{ifthen}
\usepackage{color}
\usepackage{float}
\usepackage{afterpage}
\usepackage{bm}
\allowdisplaybreaks

\newtheorem{theorem}{Theorem}

\newtheorem{lemma}{Lemma}

\newtheorem{problem}{Problem}

\floatstyle{ruled}
\newfloat{algorithm}{htb}{loa}
\floatname{algorithm}{Algorithm}

\newcommand{\beq}{\begin{equation}}
\newcommand{\eeq}{\end{equation}}
\newcommand{\ba}{\begin{array}}
\newcommand{\ea}{\end{array}}
\newcommand{\bea}{\begin{eqnarray}}
\newcommand{\eea}{\end{eqnarray}}
\newcommand{\beba}{\begin{equation}\begin{array}{lllll}}
\newcommand{\eeea}{\end{array}\end{equation}}
\newcommand{\bc}{\begin{cases}}
\newcommand{\ec}{\end{cases}}
\newcommand{\bpm}{\begin{pmatrix}}
\newcommand{\epm}{\end{pmatrix}}
\newcommand{\ben}{\begin{enumerate}}
\newcommand{\een}{\end{enumerate}}
\newcommand{\bit}{\begin{itemize}}
\newcommand{\eit}{\end{itemize}}

\newcommand{\mrm}{\mathrm}
\newcommand{\mbb}{\mathbb}
\newcommand{\mcal}{\mathcal}
\newcommand{\mbf}{\mathbf}

\newcommand{\tbf}{\textbf}

\newcommand{\tsf}{\textsf}

\newcommand{\vphi}{\varphi}

\newcommand{\bbC}{\mbb{C}}

\newcommand{\bbR}{\mbb{R}}

\newcommand{\bbU}{\mbb{U}}

\let\originalleft\left
\let\originalright\right
\renewcommand{\left}{\mathopen{}\mathclose\bgroup\originalleft}
\renewcommand{\right}{\aftergroup\egroup\originalright}

\newcommand{\lb}{\left(}
\newcommand{\rb}{\right)}
\newcommand{\lr}[1]{\lb #1 \rb}

\newcommand{\intset}[1]{\left \{1, 2, \dots, #1 \right \}}
\newcommand{\seta}[1]{\left \{ #1 \right \}}

\renewcommand{\emptyset}{\O}

\newcommand{\defeq}{\vcentcolon=}

\newcommand{\mmax}[1]{\mrm{max}\left \{ #1 \right \}}
\newcommand{\mmin}[1]{\mrm{min}\left \{ #1 \right \}}

\newcommand{\argmin}{\mathop{\operatorname{argmin}}}

\newcommand{\dsum}{\displaystyle\sum\limits}

\newcommand{\abs}[1]{\left | #1 \right |}
\newcommand{\sgn}[1]{\operatorname{sgn}\lb #1 \rb}

\newcommand{\range}[1]{\operatorname{Range}\lb #1 \rb}
\newcommand{\tr}[1]{\operatorname{tr} \lb #1 \rb}

\newcommand{\floor}[1]{\lfloor #1 \rfloor}

\newcommand{\nm}[1]{\left \| #1  \right \|}

\newcommand{\infnm}[1]{\|#1\|_{\mrm{\infty}}}
\newcommand{\frobnm}[1]{\left \| #1 \right \|_{\mrm{F}}}

\newcommand{{\matle}}{\preccurlyeq}
\newcommand{{\matge}}{\succcurlyeq}

\newcommand{\poly}[1]{\operatorname{poly}\lb #1 \rb}
\newcommand{\bgo}[1]{O \lb #1 \rb}
\newcommand{\omg}[1]{\Omega\lb #1 \rb}
\newcommand{\tht}[1]{\Theta\lb #1 \rb}
\newcommand{\lto}[1]{o \lb #1 \rb}

\renewcommand{\log}[1]{\operatorname{log} \lb #1 \rb}

\newcommand{\ket}[1]{\left | #1\right\rangle}
\newcommand{\bra}[1]{\left\langle #1\right|}
\newcommand{\braket}[2]{\left\langle #1|#2 \right\rangle}
\newcommand{\ketbra}[2]{\left|#1\rangle \langle #2  \right |}

\newcommand{\zo}{\{0,1\}}

\newcommand{\PSPACE}{\tsf{PSPACE}}

\newcommand{\BQP}{\tsf{BQP}}

\newcommand{\xor}{\oplus}

\newcommand{\proj}[1]{\mrm{\Pi}\lb #1 \rb}

\newcommand{\cond}[1]{\kappa \lb #1 \rb}

\newcommand{\bA}{\mbf{A}}

\newcommand{\bX}{\mbf{X}}

\newcommand{\bfa}{\mbf{a}}
\newcommand{\bx}{\mbf{x}}
\newcommand{\by}{\mbf{y}}
\newcommand{\bz}{\mbf{z}}
\newcommand{\bb}{\mbf{b}}
\newcommand{\bu}{\mbf{u}}
\newcommand{\bv}{\mbf{v}}

\newcommand{\beps}{\mbf{\epsilon}}

\newcommand{\px}{\mcal{P}_x}
\newcommand{\py}{\mcal{P}_y}

\begin{document}

\preprint{APS/123-QED}

\title{Quantum Algorithm for Linear Regression}
\author{Guoming Wang}
\email{wgmcreate@berkeley.edu}
\affiliation{Joint Center for Quantum Information and Computer Science, University of Maryland, College Park, MD 20742, USA} 
\date{\today}

\begin{abstract}
We present a quantum algorithm for fitting a linear regression model to a given data set using the least squares approach. Different from previous algorithms which yield a quantum state encoding the optimal parameters, our algorithm outputs these numbers in the classical form. So by running it once, one completely determines the fitted model and then can use it to make predictions on new data at little cost. Moreover, our algorithm works in the standard oracle model, and can handle data sets with nonsparse design matrices. It runs in time $\operatorname{poly}(\operatorname{log}(N), d, \kappa, 1/\epsilon)$, where $N$ is the size of the data set, $d$ is the number of adjustable parameters, $\kappa$ is the condition number of the design matrix, and $\epsilon$ is the desired precision in the output. We also show that the polynomial dependence on $d$ and $\kappa$ is necessary. Thus, our algorithm cannot be significantly improved. Furthermore, we also give a quantum algorithm that estimates the quality of the least-squares fit (without computing its parameters explicitly). This algorithm runs faster than the one for finding this fit, and can be used to check whether the given data set qualifies for linear regression in the first place.
\end{abstract}

\maketitle

\section{Introduction}
\emph{Curve fitting}, also known as \emph{regression analysis} in statistics, is the process of constructing a mathematical function that has the best fit to a series of data points according to some criterion. This procedure is widely used in many scientific fields, including physics, astronomy, chemistry, biology, medicine, agriculture, geology, engineering, economics, etc. It can help us to understand the relationship among variables, to predict the unknown value of a variable from the known values of other variables, to compress data, and to aid data visualization. In practice, one often needs to fit a concise theoretical model to a huge amount of experimental data, and it is highly desirable to have an efficient algorithm for this task. 

\emph{Linear regression} is one of the most common forms of curve fitting. It assumes that the relationship between a \emph{dependent variable} (or \emph{response}) and one or more \emph{explanatory variables} (or \emph{predictors}) is linear. So it fits a function which is linear in some adjustable parameters to the given data set. These parameters are usually determined using the \emph{(ordinary) least squares} approach, which minimizes the sum of the squared deviations of the data from the model function. This optimization problem turns out to be closely related to a matrix inversion problem,  which is time-consuming for large data sets. 

With the rise of quantum computation, one naturally asks whether quantum algorithms can perform linear regression faster than their classical counterparts. Wiebe, Braun and Llyod (WBL) \cite{wiebe2012quantum} first studied this problem and answered it affirmatively. Building upon the quantum algorithm for solving linear systems of equations by Harrow, Hassidim and Lloyd (HHL) \cite{harrow2009quantum}, they developed a quantum algorithm for estimating the quality of the least-squares fit for a given data set. Under the assumption that there exist two fast procedures for specifying the nonzero entries of the \emph{design matrix} and for preparing a quantum state proportional to the \emph{response vector}, respectively (see Section \ref{sec:linearregress} for the definition of this matrix and vector), their algorithm has complexity $\poly{\log{N},s,\kappa,1/\epsilon}$, where $N$ is the size of the data set, $s$ and $\kappa$ are the sparsity and condition number of the design matrix, respectively, and $\epsilon$ is the desired precision in the output. WBL also gave an algorithm with similar complexity for preparing a quantum state approximately proportional to the optimal  parameters. Furthermore, they proposed to use statistical sampling and quantum state tomography to find a concise representation for this state. WBL's algorithms are mainly suited for data sets whose design matrices are sparse and well-conditioned.

Recently, Schuld, Sinayskiy and Petruccione (SSP) \cite{schuld2016prediction} reapproached the problem of linear regression on a quantum computer from a machine learning perspective. Building upon HHL's strategy for matrix inversion and Lloyd, Mohseni and Rebentrost (LMR)'s \emph{density matrix exponentiation} technique \cite{lloyd2014quantum}, they developed a quantum algorithm for \emph{pattern recognition}, in which one only needs to make a prediction on a new data point based on a linear regression model trained on a given data set and does not need to find this model explicitly. Their algorithm takes as input multiple copies of three quantum states encoding the design matrix of the training set, the response vector of the training set, and the new data point, respectively, and outputs a scalar value which is the predicted response for the new data point. Excluding the costs of preparing these states and assuming the design matrix is close to a low-rank matrix, this algorithm has complexity $\poly{\log{d},\kappa,1/\epsilon}$, where $d$ is the number of adjustable parameters, $\kappa$ is the condition number of the design matrix, and $\epsilon$ is the desired precision in the output. SSP's algorithm is mainly suited for data sets whose design matrices are well-conditioned and have low-rank approximations. 

Both WBL and SSP have focused on the scenario where both the size $N$ of the data set and the number $d$ of adjustable parameters are exponentially large. Thus, they do not attempt to find the optimal parameters explicitly (which is time-consuming), but only encode these parameters in a quantum state (which can be used to make predictions on new data via swap test). While this scenario is useful in some applications (e.g. estimation of the output state of a quantum device), we believe that it is equally important to consider the scenario where $d$ is much smaller than $N$. Namely, $N$ is exponentially large, but $d$ is only polynomially large. One often encounters this situation when dealing with a classical data set and wanting to compress a large amount of data into a concise model (with few parameters). Once such a model is found explicitly, one can use it to make predictions on new data at little cost. Furthermore, saving the optimal parameters is much easier than storing the quantum state encoding these numbers, as quantum resources are fragile. 

For the above reasons, in this paper, we present a new quantum algorithm for fitting a linear regression model to a given data set using the least squares approach. Our algorithm works in the standard oracle model, and outputs the optimal parameters in the classical form. It runs in time $\poly{\log{N}, d, \kappa, 1/\epsilon}$, where $N$ is the size of the data set, $d$ is the number of adjustable parameters, $\kappa$ is the condition number of the design matrix, and $\epsilon$ is the desired precision in the output. Note that the polynomial dependence on $d$ is inevitable, because simply writing down all the optimal parameters takes $\omg{d}$ time. We show that the polynomial dependence on $\kappa$ is also necessary, by proving a lower bound on the quantum query complexity of this problem. These facts imply that our algorithm cannot be significantly improved. Furthermore, we also give a quantum algorithm that estimates the quality of the least-squares fit (without computing its parameters explicitly). This algorithm runs faster than the one for finding this fit, and can be used to check whether the given data set qualifies for linear regression in the first place. 

We make use of two recent results in designing our algorithms. The first one is Low and Chuang's method for Hamiltonian simulation based on \emph{qubitization} \cite{low2016hamiltonian} and \emph{quantum signal processing} \cite{low2017optimal}. This method allows us to simulate a nonsparse Hamiltonian, provided that this Hamiltonian can be embedded into a larger unitary operator in certain way. The second one is Childs, Kothari and Somma (CKS)'s approach to matrix inversion \cite{childs2015quantum}. This approach differs from HHL's in that it does not use phase estimation, but relies on a techique for implementing a \emph{linear combination of unitaries} (LCU) and a suitable Fourier or Chebyshev series representation of the matrix inverse function. Consequently, it has exponentially better dependence on the precision than HHL's approach. We combine these results with traditional techniques (such as amplitude estimation \cite{brassard2002quantum}) to find the optimal parameters and to estimate the quality of the least-squares fit for a given data set. 

As mentioned before, WBL have suggested a sampling-based algorithm for learning the optimal parameters in Ref.~\cite{wiebe2012quantum}. Our algorithm for computing the optimal parameters differs from their algorithm in several ways. First, as mentioned above, our algorithm uses the approach of Ref.~\cite{childs2015quantum} for matrix inversion, which has better dependence on the desired precision in the output than HHL's approach (which was used by Ref.~\cite{wiebe2012quantum}). Second, we compute the pseudoinverse of the design matrix by considering its singular value decomposition (SVD), while Ref.~\cite{wiebe2012quantum} achieved this by following a step-by-step approach (see the end of Section \ref{sec:lrfit} for more discussion on this). Third, as mentioned above, our algorithm is based on the method of Ref.~\cite{low2016hamiltonian} for simulating a large class of Hamiltonians, while Ref.~\cite{wiebe2012quantum} was based on an old method for simulating sparse Hamiltonians. As a consequence, our algorithm can handle data sets with nonsparse design matrices. Fourth, we assume that the data set is given via standard oracles (see Section \ref{sec:problem} for more details), and explicitly address the issue of preparing a quantum state proportional to the response vector. By contrast, Ref.~\cite{wiebe2012quantum} ignored the cost of this step. Finally, our algorithm uses amplitude estimation to estimate the optimal parameters, which has quadratically better dependence on the desired accuracy in the output than statistical sampling (which was used by Ref.~\cite{wiebe2012quantum}). 

The remainder of this paper is organized as follows. In Section \ref{sec:prelim}, we provide some requisite background information, and formally state the problems studied in this work. In Section \ref{sec:hs}, we describe an efficient procedure for simulating a nonsparse Hamiltonian related to the design matrix, which is a key component of our algorithms. In Section \ref{sec:lrfit}, we present a quantum algorithm for fitting a linear regression model to a given data set using the least squares approach. In Section \ref{sec:lrtest}, we propose a quantum algorithm that estimates the quality of the least-squares fit (without computing its parameters explicitly). In Section \ref{sec:lblr}, we prove a lower bound on the quantum query complexity of linear regression. Finally, we conclude in Section \ref{sec:discuss} with some comments and future research directions. 

\section{Preliminaries}
\label{sec:prelim}
In this section, we provide the necessary background information to understand this paper. In Section \ref{sec:notation}, we introduce the notation used in this paper. In Section \ref{sec:linearregress}, we review some basic facts about linear regression. In Section \ref{sec:problem}, we formally state the problems studied in this work.

\subsection{Notation}
\label{sec:notation}
Given a real number $x$, we define its sign as $\sgn{x}=1$ if $x \ge 0$, and $\sgn{x}=-1$ otherwise. Given two real numbers $a$, $b$ and a real number $\delta>0$, we say that $a$ is a $\delta$-additive approximation of $b$ if $\abs{a-b}\le \delta$. Moreover, we say that an algorithm estimates a quantity $x$ up to additive error $\delta$ if it outputs a $\delta$-additive approximation of $x$.

Given a vector $\bx=(x_1,x_2,\dots,x_N)^T \in \bbC^N$, we use $\infnm{\bx}$ and $\nm{\bx}$ to denote the $l^\infty$ and $l^2$ norms of $\bx$, respectively, i.e. 
\bea
\infnm{\bx}\defeq \max\limits_{1 \le i \le N}\abs{x_i},
\eea
 and
\bea
\nm{\bx} \defeq \sqrt{\dsum_{i=1}^{N} \abs{x_i}^2}.
\eea
Moreover, we define
\bea
\rho(\bx) \defeq \dfrac{\sqrt{N} {\infnm{\bx}}}{\nm{\bx}}
=\dfrac{\max\limits_{1 \le i \le N} \abs{x_i}}{\sqrt{\frac{1}{N} \sum\limits_{i=1}^N \abs{x_i}^2}}.
\eea
The smaller $\rho(\bx)$ is, the more \emph{balanced} $\bx$ is, in the sense that the no entry of $\bx$ has significantly larger norm than the quadratic mean norm of $\bx$'s entries. In particular, we say that $\bx$ is \emph{balanced} if $\rho(\bx)=\bgo{1}$ (e.g. at most $100$). 

Given a matrix $\bA=(a_{i,j})  \in \bbC^{N \times M}$, we define $\bfa_i \defeq (a_{i,1}, a_{i, 2},\dots, a_{i, M})^T$, for each $i \in \intset{N}$. We also
use $\nm{\bA}$ and $\frobnm{\bA}$ to denote the spectral and Frobenius norms of $\bA$, respectively, i.e. 
\bea
\nm{\bA} \defeq \max\limits_{\bx \in \bbC^M, ~\bx \neq 0} \frac{\nm{\bA \bx}}{ \nm{\bx}},
\eea
and 
\bea
\frobnm{\bA} \defeq  \sqrt{\sum_{i=1}^N \sum_{j=1}^M \abs{a_{i,j}}^2}.
\eea
 In addition, we define
\bea
\nm{\bA}_{2,\infty} &\defeq & \max\limits_{\bx \in \bbC^M, ~\bx \neq 0} \dfrac{\infnm{\bA\bx}}{\nm{\bx}}  \\
&= &\max\limits_{1 \le i \le N} \nm{\bfa_i} \\
&= &\max\limits_{1 \le i \le N} \sqrt{ \dsum_{j=1}^M \abs{a_{i, j}}^2 }
\eea
and 
\bea
\sigma(\bA) \defeq \dfrac{\sqrt{N} \nm{\bA}_{2,\infty}}{\frobnm{\bA}}
=\dfrac{\max\limits_{1 \le i\le N} \nm{\bfa_i}}{\sqrt{\frac{1}{N} \sum\limits_{i=1}^N \nm{\bfa_i}^2}}.
\eea
The smaller $\sigma(\bA)$ is, the more \emph{balanced} $\bA$ is, in the sense that no row of $\bA$ has significantly larger norm than the quadratic mean norm of $\bA$'s rows. In particular, we say that $\bA$ is \emph{balanced} if $\sigma(\bA)=\bgo{1}$ (e.g. at most $100$). 

For the above $\bA$, we also use $\range{\bA}$ to denote the range (i.e. column space) of $\bA$, and use $\proj{\bA}$ to denote the projection onto $\range{\bA}$. We also use $s_j(A)$ to denote $j$-th smallest singular value of $A$ (counted with multiplicity), and use $\lambda_j(A)$ to denote the $j$-th smallest eigenvalue of $A$ (counted with multiplicity), starting with $j=1$. The \emph{condition number} of $\bA$, denoted by $\cond{\bA}$, is defined as the ratio of largest to smallest singular value of $\bA$. Futhermore, we use $\bA^+$ to denote the \emph{Moore-Penrose pseudoinverse} of $\bA$. That is, if $\bA$ has the singular value decomposition $\bA=\sum_k s_k \bu_k \bv_k^{\dagger}$, where $s_k>0$, $\bu_k \in \bbC^N$ and $\bv_k \in \bbC^M$ are unit vectors, then $\bA^+ \defeq \sum_k s_k^{-1} \bv_k \bu_k^{\dagger}$. 

Given a matrix $\bA \in \bbC^{N \times M}$ and a vector $\bx \in \bbC^N$, we define
\bea
\tau(\bA, \bx) \defeq \dfrac{\nm{\proj{\bA} \bx}^2}{\nm{\bx}^2}.
\eea
In words, $\tau(\bA, \bx)$ measures how much ``fraction'' of $\bx$ lies in the range of $\bA$. In particular, we say that $(\bA, \bx)$ is \emph{well-behaved} if $\tau(\bA, \bx)=\omg{1}$ (e.g. at least $2/3$).

Given a vector $\bx \in \bbC^N$, we say that $\bx$ is $d$-sparse if it contains at most $d$ nonzero entries. Given a matrix $\bA  \in \bbC^{N \times M}$, we say that $\bA$ is $d$-sparse if it contains at most $d$ nonzero entries in each row and column. In particular, if $d=\poly{\log{L}}$ where $L=\mmax{N,M}$, then we simply say that $\bA$ is sparse.

Given a state $\ket{\vphi} \in \bbC^d$ and a real number $\epsilon>0$, we say that a procedure prepares $\ket{\vphi}$ with precision $\epsilon$ if this procedure prepares a state $\ket{\psi} \in \bbC^d$ satisfying $\nm{\ket{\vphi}-\ket{\psi}} \le \epsilon$. 

Given a unitary operation $V \in \bbU(d)$ and a real number $\epsilon>0$, we say that a procedure implements $V$ with precision $\epsilon$ and failure probability $\bgo{\epsilon}$ if there exists an integer $l\ge 0$ such that, on any input state $\ket{\psi} \in \bbC(d)$, this procedure first appends an $l$-qubit ancilla system in state $|0^l\rangle$, then performs a unitary operation $U \in \bbU(2^l \times d)$ on the joint  system such that
\beq
U\ket{0^l}\ket{\psi} = \ket{0^l} A \ket{\psi}
+\sum_{j \neq 0^l} \ket{j} B_j \ket{\psi},
\eeq
where $A$ and the $B_j$'s are linear operators satisfying $\nm{A-V} \le \epsilon$ and $A^{\dagger}A+\sum_{j \neq 0^l} B_j^{\dagger}B_j=I$, and finally measures the ancilla system and postselects on the outcome being $0^l$. Note that since $V$ is unitary and $\nm{V-A}\le \epsilon$, we get $\nm{V\ket{\psi} - A\ket{\psi}} \le \epsilon$, and
\bea
\abs{\nm{A\ket{\psi}} - 1}
&=&\abs{\nm{A\ket{\psi}} - \nm{V\ket{\psi}}} \\
&\le & \nm{(A-V)\ket{\psi}} \\
&\le & \epsilon,
\eea
and
\bea
\nm{V\ket{\psi} - \dfrac{A\ket{\psi}}{\nm{A\ket{\psi}}}} &\le & 
\nm{V\ket{\psi} - A\ket{\psi}}\nonumber\\
&&+
\nm{A\ket{\psi} - \dfrac{A\ket{\psi}}{\nm{A\ket{\psi}}}} \\
&\le & \epsilon + \abs{\nm{A\ket{\psi}} - 1}\\
&\le &2\epsilon.
\eea
Thus, on any input state $\ket{\psi}$, this procedure succeeds with probability $\nm{A\ket{\psi}}^2=1-\bgo{\epsilon}$ (with a flag indicating success), and when it succeeds, it outputs the state $\frac{A\ket{\psi}}{\nm{A\ket{\psi}}}$ which is $\bgo{\epsilon}$-close to $V\ket{\psi}$ in $l^2$ norm.

Now consider a quantum circuit consisting of a sequence of unitary operations $V_1 \to V_2 \to \dots \to V_{m-1} \to V_m$. Suppose $P_i$ is a procedure that implements $V_i$ with precision $\epsilon_i$ and failure probability $\bgo{\epsilon_i}$, for each $i \in \intset{m}$. Let $P$ be the concatenation of these procedures (i.e. $P_1\to P_2 \to \dots \to P_{m-1} \to P_m$). Then by a standard hybrid argument, one can show that $P$ implements the unitary operation $V\defeq V_mV_{m-1}\dots V_2V_1$ with precision $\epsilon\defeq \sum_{i=1}^m \epsilon_i$ and failure probability $\bgo{\epsilon}$. Thus, on any input state $\ket{\psi}$, the procedure $P$ succeeds with probability $1-\bgo{\epsilon}$ (with a flag indicating success), and when it succeeds, it outputs a state $\bgo{\epsilon}$-close to $V\ket{\psi}$ in $l^2$ norm. This fact will be useful in the design of our algorithms. 

\subsection{Linear Regression}
\label{sec:linearregress}
Given a data set $\seta{y_i, x_{i, 1}, x_{i, 2}, \dots, x_{i, d}}_{i=1}^N$ of $N$ statistical units (where $N \ge d$), a linear regression model assumes that the relationship between the \emph{response} (or \emph{regressand},  \emph{dependent variable}) $y_i$ and the \emph{predictors} (or \emph{regressors}, \emph{explanatory variables}) $x_{i, 1}, x_{i, 2}, \dots, x_{i, d}$ is linear. That is, there exist some unknown parameters $\beta_1$, $\beta_2$, $\dots$, $\beta_d$ and residual terms $\epsilon_i$ such that
\beq
y_i = \beta_1 x_{i, 1} + \beta_2 x_{i,2} + \dots + \beta_d x_{i,d} + \epsilon_i,  ~1 \le i \le N.
\eeq
In the matrix form, it can be written as
\bea
\by = \bX \beta + \beps,
\eea
where 
\bea
\bX \defeq
\bpm
\bx_1^T \\
\bx_2^T \\
\vdots \\
\bx_N^T
\epm
= \bpm
x_{1,1}  & x_{1,2} & \dots & x_{1,d}\\
x_{2,1}  & x_{2,2} & \dots & x_{2,d}\\
\vdots   & \vdots  &  \ddots & \vdots \\
x_{N,1}  & x_{N,2} & \dots & x_{N,d}
\epm,~~ \\
\by \defeq 
\bpm
y_1\\
y_2\\
\vdots\\
y_N
\epm, 
~~\beta \defeq 
\bpm
\beta_1 \\
\beta_2 \\
\vdots \\
\beta_d
\epm,
~~
\beps \defeq \bpm 
\epsilon_1\\
\epsilon_2\\
\vdots \\
\epsilon_N
\epm.~~~
\eea
We usually call $\bX$ the \emph{design matrix}, $\by$ the \emph{response vector}, $\beta$ the \emph{parameter vector}, and $\beps$ the \emph{residual vector}. Here we assume that the $x_{i,j}$'s and $y_i$'s are real numbers. This is actually without loss of generality, because any linear regression model with complex variables can be reduced to a (slightly larger) linear regression model with real variables. Moreover, we assume that the design matrix $\bX$ has full rank $d$. In other words, the $d$ columns of $\bX$ are linearly independent. This is a necessary condition for linear regression to have a unique solution.

We emphasize that the predictors can be nonlinear functions of some ``baseline" variables. This allows linear regression to fit a nonlinear relationship between the response and the baseline variables. For example, suppose we are interested in learning how the yield $y_i$ of a chemical synthesis is related to the temperature $t_i$ at which the synthesis takes place. We propose a quadratic model of the form:
\beq
y_i = a_0 + a_1 t_i + a_2 t_i^2 + \epsilon_i,  ~1 \le i \le N.
\eeq
This model is linear in the parameters $a_0$, $a_1$ and $a_2$, but nonlinear in the baseline variable $t_i$. In the matrix form, it can be written as
\bea
\bpm
y_1 \\
y_2 \\
\vdots\\
y_N
\epm = \bpm
1 & t_1 & t_1^2 \\
1 & t_2 & t_2^2 \\
\vdots & \vdots & \vdots \\
1 & t_N & t_N^2
\epm 
\bpm
a_0 \\
a_1 \\
a_2
\epm
+
\bpm
\epsilon_1 \\
\epsilon_2 \\
\vdots \\
\epsilon_N
\epm
\eea
Here the design matrix is a Vandermonde matrix, and it has full rank as long as there are at least three distinct $t_i$'s. Furthermore, this design matrix is not sparse. This is a generic phenomenon in linear regression, because we often include the constant $1$ as one of the predictors, and consequently the design matrix often contains a dense column of all $1$'s.

Linear regression models are usually fitted using the \emph{least squares} approach, which minimizes the sum of the squared residuals. Namely, it finds
\bea
\hat{\beta} \defeq \argmin\limits_{\beta \in \bbR^d} ~\nm{\bX \beta - \by}^2
\eea
This optimization problem has the following closed-form solution \cite{golub1970singular}
\bea
\hat{\beta} = \bX^+ \by
=(\bX^T \bX)^{-1}\bX^T \by.
\label{eq:lrsol}
\eea
Noting that 
\bea
\proj{\bX} = \bX (\bX^T \bX)^{-1}\bX^T,
\eea
we obtain 
\bea
\bX \hat{\beta} = \proj{\bX} \by.
\label{eq:lrproj}
\eea
Namely, $\bX \hat{\beta}$ is exactly the projection of $\by$ onto the range
of $\bX$. This is the geometric interpretation of least-squares linear regression.
 
Although Eq.~(\ref{eq:lrsol}) gives the solution of linear regression, it is not computationally convenient, because $\bX$ is a rectangular matrix and $\bX^+$ is not easy to implement physically. To overcome this issue, we adopt the strategy of Ref. \cite{harrow2009quantum} (which was also used in Ref. \cite{wiebe2012quantum}) and embed $\hat{\beta}$ into the solution of a larger linear system. Specifically, let 
\bea
\bA \defeq 
\bpm
0 & \bX \\
\bX^T & 0
\epm, &~
\bb
\defeq
\bpm
\by \\
0
\epm, &~
\bz \defeq
\bpm
0 \\
\hat{\beta}
\epm.
\label{eq:defabz}
\eea
Then we have
\bea
\bA^+ \bb
&=&
\bpm
0 & \bX \\
\bX^T & 0
\epm^+ \bpm
\by \\
0
\epm \\
&=&
\bpm
0 & (\bX^T)^+\\
\bX^+ & 0
\epm
\bpm
\by\\
0
\epm\\
&=&
\bpm
0 \\
\bX^+ \by
\epm \\
&=& \bpm
0 \\
\hat{\beta}\\
\epm\\
&=& \bz.
\label{eq:foundation}
\eea
The fact that $\bA$ is a real symmetric matrix facilitates the implementation of $\bA^+$. Once we have a procedure for preparing a quantum state proportional to $\bA^+\bb=\bz$, we can utilize this procedure to get useful information about $\hat{\beta}$. 

A statistical model fits a data set well only if the discrepancy between the observed response and the response predicted by this model is small. Here we measure the quality of the least-squares fit $\by \approx \bX \hat{\beta}$ using the quantity
\bea
\tau \defeq \dfrac{\nm{\hat{\by}}^2}{\nm{\by}^2}
=1 - \dfrac{\nm{\hat{\beps}}^2}{\nm{\by}^2},
\eea
where 
\bea
\hat{\by} \defeq \bX \hat{\beta} = \proj{\bX} \by
\eea
and 
\bea
\hat{\beps} \defeq \by-\hat{\by}=(I- \proj{\bX}) \by.
\eea
Namely, $1-\tau$ is the ratio of the squared norm of the residual vector $\hat{\beps}$ to that of the response vector $\by$. It turns out that $\tau = \tau(\bX,\by)$. So a data set $\lr{\bX, \by}$ can be explained well by a linear regression model only if it is well-behaved, i.e. $\tau(\bX, \by) = \omg{1}$ (i.e. at least $2/3$). This kind of data sets will be the main focus of our study. We will also give an efficient quantum algorithm for testing whether a given data set is well-behaved or not. 

It is worth noting that Wiebe, Braun and Lloyd (WBL) \cite{wiebe2012quantum} have used the quantity $E=\nm{\ket{\by}-\mbf{F}\ket{\bm{\lambda}}}^2$ (according to their notation) to measure the error of the least-squares fit. Their $\ket{\by}$, $\mbf{F}$ and $\ket{\bm{\lambda}}$ correspond to our $\by/\nm{\by}$, $\bX$ and $\hat{\beta}/\nm{\by}$, respectively. Then one can see that their $1-E$ is equivalent to our $\tau$. So WBL essentially measured the quality of the least-squares fit in the same way as we do. 

In practice, after one collects the raw data from the experiements, one does not immediately fit a mathematical function to these data. Instead, one needs to preprocess the raw data to make them well-suited for data fitting. This preprocessing usually consists of imputation of missing data, data normalization or standadization, and elimination of influential \emph{outliers} which have detrimental effect on the estimated regression function (an outlier is a data point whose response $y$ does not follow the general trend of the rest of the data). The last step is important, because we want the fitted model to capture the \emph{typical} relationship among the response and predictors so that it can be generalized to new data. This requires that the loss function 
\bea
\nm{\bX \beta - \by}^2 &=& 
\dsum_{i=1}^N \abs{\bx_i^T \beta - y_i}^2 \\
&=&\dsum_{i=1}^N \abs{\lb \dsum_{j=1}^d \beta_j x_{i,j} \rb - y_i}^2,
\eea
should not be dominated by only a few data points. An outlier has the potential to do so, especially if it has high \emph{leverage} (i.e. it has ``extreme" predictor $x$ values). However, we emphasize that not all outliers are influential and should be eliminated. The identification of influential outliers is an important and complicated topic, and many techniques have been developed for this task, such as difference in fits (DFFITS) and Cook's distance. In this paper, we assume that the given data set has already been preprocessed, and the harmful outliers have been removed. Since there is no general characterization of such data points, we assume for simplicity that no $\bx_i$ or $y_i$ has extremely large norm (compared to the average norm of the $\bx_i$'s or $y_i$'s, respectively). In other words, both $\bX$ and $\by$ are balanced, i.e. $\sigma(\bX)=\bgo{1}$ (e.g. at most $100$) and $\rho(\by)=\bgo{1}$ (e.g. at most $100$). These assumptions ensure that no data point has  significantly larger contribution to the loss function than the others, and are useful in practice. However, we acknowledge that these assumptions might be too stringent for some applications, and standard preprocessing techniques do not always guarantee them, and it remains future work to extend our results to the most general case of linear regression. 

\subsection{Problem Statement}
\label{sec:problem}

In this paper, we assume that the data set $\seta{y_i, x_{i,1},x_{i,2},\dots,x_{i,d}}_{i=1}^N$ is given via two black-box subroutines. For $\bX=(x_{i,j}) \in \bbR^{N \times d}$, we assume there exists a procedure $\px$ that allows us to perform the map
\beq
\ket{i}\ket{j}\ket{z} \mapsto \ket{i}\ket{j}\ket{z\xor x_{i,j}}
\eeq
for any $i \in \intset{N}$ and $j \in \intset{d}$, where the third register holds a bit string representing an entry of $\bX$. For $\by=(y_1,y_2,\dots,y_N)^T \in \bbR^N$, we assume there exists a procedure $\py$ that allows us to perform the map
\beq
\ket{i}\ket{z} \mapsto \ket{i}\ket{z\xor y_{i}}
\eeq
for any $i \in \intset{N}$, where the second register holds a bit string representing an entry of $\by$. We assume that both $\px$ and $\py$ are efficient, in the sense that they run in time $\poly{\log{N}}$. This requires that either each entry of $\bX$ and $\by$ can be quickly computed by an algorithm (given its position), or these entries are stored in a quantum random access memory (QRAM) beforehand. Our algorithms work well in both cases. 

Given access to $\px$ and $\py$, our primary goal is to fit a linear regression model to the data set $\seta{y_i, x_{i,1},x_{i,2},\dots,x_{i,d}}_{i=1}^N$ using the least squares approach. Our secondary goal is to estimate the quality of the fitted model (without computing its parameters explicitly). 

Formally, we define our linear regression (LR) problems as follows:
\begin{problem}[LR-P]
Let $\bX=(x_{i,j}) \in \bbR^{N \times d}$ be a balanced matrix such that its singular values are in the range $[1/\kappa, 1]$.  Let $\by=(y_1,y_2,\dots,y_N)^T \in \bbR^N$ be a balanced unit vector. Suppose $(\bX, \by)$ is well-behaved. Given $\epsilon>0$ and access to the procedures $\px$ and $\py$ described above, the goal is to output a vector $\beta \defeq (\beta_1, \beta_2, \dots, \beta_d)^T \in \bbR^d$ satisfying $\infnm{\beta-\hat{\beta}} \le \epsilon$, where $\hat{\beta} \defeq \bX^+ \by$, succeeding with high probability (e.g. at least $2/3$). 
\label{prob:1}
\end{problem}

\begin{problem}[LR-Q]
Let $\bX=(x_{i,j}) \in \bbR^{N \times d}$ be a balanced matrix such that its singular values are in the range $[1/\kappa, 1]$. Let $\by=(y_1,y_2,\dots,y_N)^T \in \bbR^N$ be a balanced unit vector. Given $\epsilon>0$ and access to the procedures $\px$ and $\py$ described above, the goal is output an $\epsilon$-additive approximation of $\tau \defeq {\nm{\proj{\bX} \by}^2}/{\nm{\by}^2}$, succeeding with high probability (e.g. at least $2/3$). 
\label{prob:2} 
\end{problem}

Although in the above problems we assume that the singular values of $\bX$ lie in the range $[1/\kappa, 1]$ and $\nm{\by}=1$, this is without loss of generality. Suppose instead that the singular values of $\bX$ lie in the range $[a/\kappa, a]$ and $\nm{\by}=b$, for some constants $a, b>0$. Namely, $\bX$ and $\by$ are rescaled by a factor of $a$ and $b$, respectively. Then $\hat{\beta}=\bX^+ \by$ is rescaled by a factor of $b/a$. So we only need to multiply the result of LR-P by this factor. On the other hand, $\tau={\nm{\proj{\bX} \by}^2}/{\nm{\by}^2}$ is immune to this rescaling. So we do not need to make any change to the result of LR-Q.

We will develop quantum algorithms for solving the above problems. We quantify the resource requirements of these algorithms using two measures. The \emph{query complexity} is the number of uses of the procedures $\px$ and $\py$ in the algorithm. The \emph{gate complexity} is the number of 2-qubit gates used in the algorithm. An algorithm is \emph{gate-efficient} is if it is gate complexity is larger than its query complexity only by a logarithmic factor. Formally, an algorithm with query complexity $Q$ is gate-efficient if its gate complexity is $\bgo{Q \cdot \poly{\log{QN}}}$. All the algorithms presented in this paper will be gate-efficient.

\section{Hamiltonian Simulation}
\label{sec:hs}

Hamiltonian simulation is an important topic that has received a lot of attention in the past years \cite{lloyd1996universal,aharonov2003adiabatic,childs2004quantum,berry2007efficient,
childs2010on,childs2011simulating,poulin2011quantum,childs2012hamiltonian,berry2012blackbox,berry2014exponential,berry2015simulating,berry2015hamiltonian,
low2016hamiltonian,low2017optimal}. Recently, Low and Chuang \cite{low2017optimal} proposed a techinque named \emph{quantum signal processing}, and showed how to use this technique and Childs' quantum walk \cite{childs2010on} to simulate sparse Hamiltonians nearly optimally. Later, 
in Ref.~\cite{low2016hamiltonian}, they proposed another technique called  \emph{qubitization}, and demonstrated how to use this technique and quantum signal processing to simulate a larger class of Hamiltonians efficiently. In this paper, we will use the following variant of their result:

\begin{theorem}[Adapted from Theorem 1 of Ref.~\cite{ low2016hamiltonian}]
Let $\hat{U}$ and $\hat{G}$ be unitary operators on $n$ and $k ~(<n)$ qubits,  respectively, such that $\bra{G}\hat{U}\ket{G}=\hat{H}$ is a Hermitian operator on $n-k$ qubits, where $\ket{G} \defeq \hat{G}\ket{0^k}$. Then there exists a gate-efficient algorithm that simulates $e^{-i\hat{H}t}$ with precision $\epsilon$ and failure probability $\bgo{\epsilon}$ by making $\bgo{t+\log{1/\epsilon}}$ uses of controlled-$\hat{G}$ and controlled-$\hat{U}$.
\label{thm:hsqubit}
\end{theorem}

Theorem \ref{thm:hsqubit} provides a way to simulate a nonsparse Hamiltonian, provided that this Hamiltonian can be embedded into a larger unitary operator in the way described above. Using this fact, we develop an efficient procedure for simulating $e^{-i\bA t}$, which will be a crucial component of our algorithms for solving the LR-P and LR-Q problems. Recall that $\bA$ is defined by Eq.~(\ref{eq:defabz}) and is not sparse in general.

\begin{lemma}
Let $\bX$ be defined as in LR-P or LR-Q. Let $\bA \defeq \ketbra{1}{0}\otimes \bX^T+\ketbra{0}{1}\otimes \bX$. Then there exists a gate-efficient procedure that simulates $e^{-i \bA t}$ with precision $\epsilon$ and failure probability $\bgo{\epsilon}$ by making $$\bgo{ d  \lb \sqrt{d} t+ \log{\dfrac{1}{\epsilon}} \rb }$$ uses of $\px$.
\label{lem:hs}
\end{lemma}
\begin{proof}
Let $\tilde{\bX}=(\tilde{x}_{i,j}) \defeq \bX \cdot \sqrt{N} / (\sigma \sqrt{d})$, where $\sigma \defeq \sigma(\bX) = \bgo{1}$. Then we claim 
\bea
\nm{\tilde{\bX}}_{2,\infty} 
&=&\max\limits_{1 \le i \le N} \nm{\tilde{\bx}_i} \\
&=&\max\limits_{1 \le i \le N} \sqrt{\dsum_{j=1}^d \abs{\tilde{x}_{i,j} }^2}\\ 
&\le & 1.
\label{eq:tildebxnm}
\eea
To see this, recall that the singular values of ${\bX}$ are in the range $[1/\kappa, 1]$. So
\bea
\frobnm{\bX}^2 
= \tr{\bX^T \bX} = \dsum_{j=1}^d (s_j(\bX))^2 \le d.
\eea
This implies that
\bea
\nm{\bX}_{2,\infty}
&=&\max\limits_{1 \le i \le N} \nm{\bx_i}\\
&=&\dfrac{\sigma \frobnm{\bX}}{\sqrt{N}}\\
&\le & \dfrac{\sigma \sqrt{d}}{\sqrt{N}}.
\label{eq:nmxibound}
\eea
Using this fact and $\tilde{\bX}=\bX \cdot \sqrt{N} / (\sigma \sqrt{d})$, we obtain Eq.~(\ref{eq:tildebxnm}), as desired.

Now let $\hat{V}$ be a unitary operator such that
\beq
\hat{V} \ket{0, i}_1 \ket{0, 0^m}_2 \ket{0}_3 
= \ket{0, i}_1 \ket{\vphi_i}_{2,3},~1 \le i \le N,~~~
\label{eq:defv11}
\eeq
\beq
\hat{V} \ket{1, j}_1 \ket{0, 0^m}_2 \ket{0}_3 
= \ket{1, j}_1 \ket{\psi}_{2,3}, ~1 \le j \le d,~~~
\label{eq:defv12}
\eeq
where $m=\tht{\log{N}}$, 
\beq
\ket{\vphi_i}_{2,3} \defeq  \dsum_{j=1}^d \tilde{x}_{i,j}\ket{1, j}_{2} \ket{0}_{3} +\sqrt{1-\nm{\tilde{\bx}_i}^2} \ket{1, 1}_{2} \ket{1}_{3}~~~
\label{eq:hsvstate}
\eeq
and
\bea
\ket{\psi}_{2,3} \defeq \frac{1}{\sqrt{N}} \sum_{i=1}^N \ket{0, i}_{2}\ket{0}_{3}.
\eea
Let $\operatorname{SWAP}_{1, 2}$ be the swap operator on the first two registers, i.e. $\operatorname{SWAP}_{1,2} \ket{\vphi}_1 \ket{\psi}_2 = \ket{\psi}_1 \ket{\vphi}_2$ for all states $\ket{\vphi}$ and $\ket{\psi}$. Then we define
\bea
\hat{W} \defeq  (\operatorname{SWAP}_{1,2} \otimes I_{3}) \cdot \hat{V}
\label{eq:defv2}
\eea
and 
\bea
\hat{U} \defeq \hat{W}^{\dagger} \hat{V}.
\label{eq:defu1}
\eea
In addition, let $\ket{G} \defeq \ket{0, 0^m}_2 \ket{0}_3$. Then by a direct calculation, one can verify that
\bea
\hat{H} \defeq \bra{G} \hat{U} \ket{G}  =\dfrac{\bA}{\sigma \sqrt{d}}.
\eea
We will show below that $\hat{U}$ can be implemented by a gate-efficient procedure that makes use $\bgo{d}$ uses of $\mcal{P}_x$. Then by Theorem \ref{thm:hsqubit}, 
\bea
e^{-i\bA t}=e^{-i \hat{H} \sigma \sqrt{d} t}
\eea
can be implemented with precision $\epsilon$ and failure probability $\bgo{\epsilon}$ by a gate-efficient procedure that makes 
\bea
&&\bgo{ d \lb \sigma \sqrt{d} t+ \log{\dfrac{1}{\epsilon}} \rb} \\
&=&\bgo{ d \lb \sqrt{d} t+ \log{\dfrac{1}{\epsilon}} \rb}
\eea
uses of $\mcal{P}_x$ (recall that $\sigma=\bgo{1}$), as claimed.

Clearly, $\operatorname{SWAP}_{1,2}$ can be implemented in time $\poly{\log{N}}$. So it remains to show that $\hat{V}$ can be implemented by a gate-efficient procedure that makes use $\bgo{d}$ uses of $\mcal{P}_x$. To prove this, first note that the mapping 
\bea
\ket{1,j}_1 \ket{0, 0^m}_2 \ket{0}_3
\to 
\ket{1, j}_1 \ket{\psi}_{2,3}
\eea
can be implemented in time $\bgo{{\log{N}}}$, since $\ket{\psi}_{2,3} = \frac{1}{\sqrt{N}} \sum_{i=1}^N \ket{0, i}_{2}\ket{0}_{3}$ is easy to prepare. Meanwhile, we can accomplish the transformation
\bea
\ket{0, i}_1 \ket{0, 0^m}_2 \ket{0}_3
\to 
\ket{0, i}_1 \ket{\vphi_i}_{2,3}
\eea  
as follows. First, we learn $x_{i,1}$, $x_{i,2}$, $\dots$, $x_{i,d}$ by making $\bgo{d}$ uses of $\px$, and obtain the state
\bea
\ket{0, i}_1 \ket{0, 0^m}_2 \ket{0}_3 \lb \bigotimes\limits_{j=1}^d \ket{x_{i,j}}\rb_4.
\eea 
Then, we perform a unitary operation on the second and third registers 
depending on the content of the last register, and convert $\ket{0,0^m}_2 \ket{0}_3$ into $\ket{\vphi_i}_{2,3}$. This step can be achieved in time $\bgo{d \cdot \log{N}}$, since $\ket{\vphi_i}_{2,3}=\sum_{j=1}^d \tilde{x}_{i,j}\ket{1, j}_{2} \ket{0}_{3}
+\sqrt{1-\nm{\tilde{\bx}_i}^2} \ket{1, 1}_{2} \ket{1}_{3}$ is a $\bgo{d}$-sparse vector in an $\bgo{N}$-dimensional Hilbert space \cite{shende2006synthesis}. Finally, we uncompute the $x_{i,j}$'s in the last register by making $\bgo{d}$ uses of $\px$, and obtain the desired state $\ket{0,i}_1\ket{\vphi_i}_{2,3}$. This process requires $\bgo{d}$ uses of $\px$ and is gate-efficient. Combining the above facts, we know that $\hat{V}$ can be implemented by a gate-efficient procedure that makes $\bgo{d}$ uses of $\px$, as claimed.
\end{proof}

We remark that our embedding construction in the proof of Lemma \ref{lem:hs} looks similar to the construction in Refs.~\cite{childs2010on,berry2012blackbox}. However, we emphasize that our high-level strategy for simulating the Hamiltonian $\bA$ is very different from that of Refs.~\cite{childs2010on,berry2012blackbox}. Specifically, Refs.~\cite{childs2010on,berry2012blackbox} simulate a Hamiltonian by embedding it into a quantum walk operator and then performing phase estimation on this operator. By  contrast, we simulate $e^{-i\bA t}$ by embedding $\bA$ into a unitary operator $\hat{U}$ (which is not a quantum walk) in certain way and then invoking the method of Ref.~\cite{low2016hamiltonian} for Hamiltonian simulation (which is arguably more advanced than phase-estimation-based methods). So the similarity between the proof of Lemma \ref{lem:hs} and the construction in Refs.~\cite{childs2010on,berry2012blackbox} is superficial rather than essential. 

\section{Finding the Least-Squares Fit}
\label{sec:lrfit}

In this section, we present a quantum algorithm for solving the LR-P problem, i.e. finding the parameters $\hat{\beta}=\bX^+\by$ of the least-squares fit $\by \approx \bX \hat{\beta}$ for a given data set $(\bX, \by)$. Roughly speaking, this algorithm computes $\hat{\beta}=(\hat{\beta}_1,\hat{\beta}_2,\dots,\hat{\beta}_d)^T$ in three stages. The first stage estimates the absolute values of the $\hat{\beta}_i$'s. The second stage determines the signs of these parameters, up to a global sign $\pm 1$. That is, up to this stage, we obtain a vector $\beta \in \bbR^d$ which is close to either $\hat{\beta}$ or $-\hat{\beta}$. The final stage decides which of the two cases holds. This algorithm relies on several subroutines (besides the one for Hamiltonian simulation in Lemma \ref{lem:hs}). One is the following procedure for preparing the state $\ket{\by}=\sum_{i=1}^N y_i \ket{i}$ (recall that $\nm{\by}=1$). 

\begin{lemma}
Let $\by$ be defined as in LR-P or LR-Q. Then the state $\ket{\by}=\sum_{i=1}^N y_i \ket{i}$ can be prepared with precision $\delta$ by a gate-efficient procedure that makes $\bgo{\log{1/\delta}}$ uses of $\py$. 
\label{lem:preparey}
\end{lemma}
\begin{proof}
Consider the following procedure which transforms $\ket{0^n}$ into $\ket{\by}$  probabilistically, where $n=\tht{\log{N}}$. First, we map $\ket{0^n}$ to $\frac{1}{\sqrt{N}} \sum_{i=1}^N \ket{i}$ in time $\bgo{\log{N}}$. Then, we convert this state into $\frac{1}{\sqrt{N}} \sum_{i=1}^N \ket{i} \ket{y_i}$ by making $\bgo{1}$ uses of $\py$. Next, we append an ancilla qubit in state $\ket{0}$, and perform the controlled-rotation
\beq
\ket{y_i} \ket{0} \to \ket{y_i} \lb \dfrac{y_i}{\infnm{\by}} \ket{0} + \sqrt{1-\dfrac{\abs{y_i}^2}{\infnm{\by}^2}} \ket{1}\rb 
\eeq
on the last two registers, where $\infnm{\by}=\max_i \abs{y_i}=\tht{\frac{1}{\sqrt{N}}}$ (since $\by$ is a balanced unit vector). After that, we measure the ancilla qubit, and with probability
$\frac{\nm{\by}^2}{N\infnm{\by}^2}=\omg{1}$,
the outcome is $0$ and we obtain the state 
$\sum_{i=1}^N y_i \ket{i}\ket{y_i}$. Finally, we uncompute $y_i$ in the second register by making $\bgo{1}$ uses of $\py$, and obtain the desired state $\ket{\by}=\sum_{i=1}^N y_i \ket{i}$. 

The above procedure, denoted by $\mcal{A}$, makes $\bgo{1}$ uses of $\py$, is gate-efficient, and has success probability $\omg{1}$. We use Grover's $\pi/3$-amplitude amplification (i.e. the generalization of fixed-point quantum search) \cite{grover2005fixed} to raise the success probability to $1-\bgo{\delta^2}$. This boosted procedure, denoted by $\mcal{A}'$, requires $\bgo{\log{1/\delta}}$ repetitions of $\mcal{A}$, and satisfies
\bea
\mcal{A}' \ket{0^l} \ket{0^n}
= \sqrt{1-\delta'} \ket{0^l} \ket{\by} + \sqrt{\delta'}\ket{\Phi^{\perp}},
\eea
where $l$ is a positive integer, $\delta'=\bgo{\delta^2}$, and $\ket{\Phi^{\perp}}$ is a normalized state satisfying $(\ketbra{0^l}{0^l} \otimes I) \ket{\Phi^{\perp}}=0$. This implies that
\bea
\nm{\mcal{A}' \ket{0^l} \ket{0^n} - \ket{0^l} \ket{\by}}^2
&=&(1-\sqrt{1-\delta'})^2 + \delta' ~~~~\\
&=&\bgo{\delta^2}.~~~~
\eea
Furthermore, $\mcal{A}'$ makes $\bgo{\log{1/\delta}}$ uses of $\py$, and is gate-efficient. So $\mcal{A}'$ satisfies all the desired properties. This concludes the proof.
\end{proof}

Our algorithm for solving the LR-P problem also requires the following procedures for computing $|\hat{\beta}_i|$ and $|\hat{\beta}_i - \hat{\beta}_j|$.
 
\begin{lemma}
Let $\bX$, $\by$ and $\hat{\beta}$ be defined as in LR-P. Then there exists a gate-efficient quantum algorithm that makes
$$\bgo{\dfrac{d^{1.5}\kappa^3}{\epsilon^2} \cdot\poly{\log{\dfrac{\kappa}{\epsilon \delta}} }}$$
uses of $\px$ and $\py$, and outputs an 
$\epsilon$-additive approximation of $|{\hat{\beta}_i}|$, for any given $i \in \intset{d}$, succeeding with probability at least $1-\delta$.
\label{lem:qlssolentry}
\end{lemma}

\begin{proof}
Let $\bA \defeq \ketbra{1}{0} \otimes \bX^T + \ketbra{0}{1} \otimes \bX$ and $\ket{\bb} \defeq \ket{0}\ket{\by}$. Suppose $\bX$ has the singular value decomposition
\bea
\bX = \dsum_{j=1}^d s_j \ketbra{\bu_j}{\bv_j},
\eea
where $s_j \in [1/\kappa, 1]$, $\ket{\bu_j} \in \bbR^{N}$ and $\ket{\bv_j} \in \bbR^d$ are unit vectors, for all $j \in \intset{d}$. Then $\bA$ has the spectral decomposition
\bea
\bA = \dsum_{j=1}^d s_j  \ketbra{+_j}{+_j} - \dsum_{j=1}^d s_j  \ketbra{-_j}{-_j} ,
\eea
where 
\bea
\ket{\pm_{j}} \defeq \dfrac{1}{\sqrt{2}} \lb \ket{0}\ket{\bu_j} \pm  \ket{1} \ket{\bv_j} \rb.
\label{eq:ketpmj}
\eea
So $\bA $ is a Hermitian matrix whose nonzero eigenvalues are in the range $D_{\kappa} \defeq [-1, -1/\kappa] \cup [1/\kappa, 1]$. Moreover, $\ket{\bb}=\ket{0}\ket{\by}$ is a unit vector, and $\bA^+ \ket{\bb}=\ket{1} |\hat{\beta}\rangle$ by Eq.~(\ref{eq:foundation}). 

We will use a recent technique proposed by Childs, Kothari and Somma \cite{childs2015quantum} to approximately invert the matrix $\bA$.  Let the function $h(x)$ be defined as
\bea
h(x) \defeq \dsum_{j=0}^{J-1} \dsum_{k=-K}^{K} \alpha(j, k) e^{-i x \eta(j, k)},
\eea
where
\bea
&\alpha(j, k) \defeq \dfrac{i}{\sqrt{2\pi}} k \delta_y \delta_z^2  e^{-k^2 \delta_z^2/2}, \\ 
&\eta(j, k) \defeq j k \delta_y \delta_z,
\eea
for some $J=\tht{(\kappa/\epsilon) \cdot \log{\kappa/\epsilon}}$, $K=\tht{\kappa \cdot \log{\kappa/\epsilon}}$, $\delta_y=\tht{\epsilon/\sqrt{\log{\kappa/\epsilon}}}$ and $\delta_z=\tht{1/(\kappa \sqrt{\log{\kappa/\epsilon}})}$. Then $h(x)$ is $\epsilon$-close to $1/x$ on the domain $D_{\kappa}$ \cite{childs2015quantum}, i.e. 
\bea
\abs{h(x) - x^{-1}} \le \epsilon,  ~~~ \forall x \in D_{\kappa}.
\eea 
Then since $\bA$ is a Hermitian matrix whose nonzero eigenvalues are in the range $D_{\kappa}$, we have
\bea
\nm{h(\bA)-\bA^+} \le \epsilon.~~~
\eea
This implies that
\bea
\nm{h(\bA)\ket{\bb} - \bA^+ \ket{\bb}} \le \epsilon,
\label{eq:fourierapprox1}
\eea 
as $\ket{\bb}$ is a unit vector. Moreover, Ref.~\cite{childs2015quantum} shows that
\beq
\alpha \defeq \dsum_{j=0}^{J-1} \dsum_{k=-K}^{K} \abs{\alpha(j, k)} = \tht{\kappa \sqrt{\log{\kappa / \epsilon}}},
\label{eq:alphabound}
\eeq
and 
\beq
\abs{\eta(j, k)} \le JK\delta_y \delta_z = \tht{\kappa\cdot \log{\kappa/\epsilon}},
\label{eq:etajkbound}
\eeq
for all $j$, $k$. 

Now let $\ket{\bz} \defeq {\bA^{+}\ket{\bb}}=\ket{1}|{\hat{\beta}}\rangle$ and $\ket{\bz'} \defeq h(\bA) \ket{\bb}$. Then $\nm{\ket{\bz}-\ket{\bz'}} =\bgo{\epsilon}$ by Eq.~(\ref{eq:fourierapprox1}). Thus, for any $i \in \intset{d}$, we have
\bea
\abs{ \braket{1, i}{\bz'}
-{\hat{\beta}_i}}
&=&
\abs{\braket{1, i}{\bz'}
- \braket{1, i}{\bz}}\\
&\le& 
\nm{\ket{\bz'}-\ket{\bz}}\\
&=&\bgo{\epsilon}.
\eea
So in order to estimate $|{\hat{\beta}_i}|$ up to additive error $\bgo{\epsilon}$, we only need to obtain an $\bgo{\epsilon}$-additive approximation of $\abs{\braket{1, i}{\bz'}}$. This can be achieved as follows.

Let $V$ be a unitary operator such that
\bea
V \ket{0^m} = \dfrac{1}{\sqrt{\alpha}} \dsum_{j=0}^{J-1}\dsum_{k=-K}^{K} 
\sqrt{\abs{\alpha(j, k)}} \ket{j, k},
\eea
where $m=\bgo{\log{JK}}=\bgo{\log{\kappa/\epsilon}}$, and let $U$ be defined as
\beq
U \defeq i \dsum_{j=0}^{J-1} \dsum_{k=-K}^{K} \ketbra{j, k}{j, k} \otimes \sgn{k} e^{-i \bA \eta(j, k)}.
\label{eq:defu}
\eeq 
Then we define
\bea
W \defeq V^{\dagger} U V.
\eea
A direct calculation shows that
\bea
W \ket{0^m} \ket{\bb}
&=& \dfrac{1}{\alpha} \ket{0^m} {h(\bA) \ket{\bb}} + \ket{\Phi^{\perp}}\\
&=& \lb \dfrac{\nm{h(\bA) \ket{\bb}}}{\alpha} \rb \ket{0^m} \dfrac{h(\bA) \ket{\bb}}{\nm{h(\bA)\ket{\bb}}} \nonumber\\
&+& \ket{\Phi^{\perp}},
\label{eq:lcu2}
\eea
where $\ket{\Phi^{\perp}}$ is an unnormalized state satisfying 
$(\ketbra{0^m}{0^m} \otimes I) \ket{\Phi^{\perp}}=0$. Next, let $R$ be a unitary operator such that 
\bea
&R\ket{0}\ket{1,i} = \ket{1}\ket{1,i}, &~~~~~~~~~ \label{eq:r1}\\
&R\ket{0}\ket{1,i'}=\ket{0}\ket{1,i'}, &~1 \le i' \le N, ~ i' \neq i, \label{eq:r2}\\
&R\ket{0}\ket{0,j}=\ket{0}\ket{0,j}, &~1\le j \le N. \label{eq:r3}
\eea
Then by Eqs.~(\ref{eq:lcu2}), (\ref{eq:r1}), (\ref{eq:r2}) and (\ref{eq:r3}), we obtain
\bea
R W  \ket{0^m}_1 \ket{0}_2 \ket{\bb}_3 
&=&\dfrac{\braket{1,i}{\bz'}}{\alpha} \ket{0^m}_1 \ket{1}_2 \ket{1,i}_3 \nonumber\\
&+&\dsum_{i' \neq i} \dfrac{\braket{1,i'}{\bz'}}{\alpha} \ket{0^m}_1 \ket{0}_2 \ket{1,i'}_3 \nonumber \nonumber\\
&+&\dsum_{j} \dfrac{\braket{0,j}{\bz'}}{\alpha} \ket{0^m}_1 \ket{0}_2 \ket{0,j}_3 \nonumber\\
&+&\ket{\Xi^{\perp}}_{1,2,3},
\eea
where $W$ acts on the first and third registers, $R$ acts on the second and third registers, and $\ket{\Xi^{\perp}}$ is an unnormalized state satisfying 
$(\ket{0^m}\bra{0^m}\otimes I) \ket{\Xi^{\perp}}=0$. If we measure the first $m+1$ qubits of this state in the standard basis, the probability of getting  outcome $0^m 1$ is 
\bea
p' \defeq \dfrac{\abs{\braket{1,i}{\bz'}}^2}{\alpha^2}.
\eea
We use amplitude estimation \cite{brassard2002quantum} to obtain an $\epsilon''$-additive approximation $\hat{p}'$ of $p'$, where 
\bea
\epsilon'' \defeq \tht{\dfrac{\epsilon^2 }{\alpha^2}} =\tht{\dfrac{\epsilon^2}{\kappa^2  \log{\kappa/\epsilon}}},
\eea
succeeding with probability at least $3/4$. Then $\sqrt{\hat{p}'}$ is an $\bgo{\sqrt{\epsilon''}}$-additive approximation of
$\sqrt{p'}$ (note that $\sqrt{a}-\sqrt{\gamma} \le \sqrt{a-\gamma} \le \sqrt{a+\gamma} \le \sqrt{a}+\sqrt{\gamma}$ for all $a \ge \gamma \ge 0$). As a result,
\bea
\abs{\abs{\braket{1,i}{\bz'}} - \alpha \sqrt{\hat{p}'}}
&=&\abs{\alpha \sqrt{p'} - \alpha \sqrt{\hat{p}'}}\\
&=&\bgo{ \alpha \sqrt{\epsilon''}}  \\
&=&\bgo{\epsilon}.
\label{eq:apiy}
\eea
Namely, $\alpha \sqrt{\hat{p}'}$ is an $\bgo{\epsilon}$-additive approximation of $\abs{\braket{1, i}{\bz'}}$, as desired. 
 
The above basic algorithm has success probability at least $3/4$. To boost the success probability to at least $1-\delta$, we repeat this algorithm $\bgo{\log{1/\delta}}$ times, and take the median of the estimates from these runs. A standard Chernoff's bound ensures that the failure probability is at most $\delta$.

Let us analyze the complexity of this algorithm. Since we want to estimate $p'$ up to additive error $\epsilon''$, amplitude estimation requires 
\bea
\bgo{\dfrac{1}{\epsilon''}}=\bgo{\dfrac{\alpha^2}{\epsilon^2}}
=\bgo{\dfrac{\kappa^2 \log{\kappa/\epsilon}}{\epsilon^2}}
\eea
repetitions of $R$, $W=V^{\dagger}UV$ and the procedure for preparing $\ket{\bb}=\ket{0}\ket{\by}$. This means that we need to implement $U$ 
with precision $\bgo{\epsilon''}$ and failure probability $\bgo{\epsilon''}$. We also need to prepare $\ket{\by}$ with precision $\bgo{\epsilon''}$. By Lemma \ref{lem:hs}, Eqs.~(\ref{eq:etajkbound}) and (\ref{eq:defu}), and Lemma 8 of Ref.~\cite{childs2015quantum}, $U$ can be implemented with precision $\bgo{\epsilon''}$ and failure probability $\bgo{\epsilon''}$ by a gate-efficient procedure that makes $\bgo{d^{1.5} \kappa \cdot \poly{\log{\kappa/\epsilon}}}$ uses of $\px$. Meanwhile, by Lemma \ref{lem:preparey}, $\ket{\by}$ can be prepared with precision $\bgo{\epsilon''}$ by a gate-efficient procedure that makes $\bgo{\log{\kappa/\epsilon}}$ uses of $\py$. Furthermore, $V$ can be implemented in time $\bgo{\kappa \cdot \poly{\log{\kappa/\epsilon}}}$ \cite{childs2015quantum}, and clearly $R$ can be implemented in time $\poly{\log{N}}$. As a result, this algorithm makes 
\bea
\bgo{\dfrac{d^{1.5}\kappa^3}{\epsilon^2} \cdot \poly{\log{\dfrac{\kappa}{\epsilon\delta}} }}
\eea
uses of $\px$ and $\py$, and is gate-efficient, as claimed. 
\end{proof}

\begin{lemma}
Let $\bX$, $\by$ and $\hat{\beta}$ be defined as in LR-P. Then there exists a gate-efficient quantum algorithm that makes
$$\bgo{\dfrac{d^{1.5}\kappa^3}{\epsilon^2} \cdot\poly{\log{\dfrac{\kappa}{\epsilon \delta}} }}$$
uses of $\px$ and $\py$, and outputs an $\epsilon$-additive approximation of $|{\hat{\beta}_i-\hat{\beta}_j}|$, for any given $i, j \in \intset{d}$, succeeding with probability at least $1-\delta$.
\label{lem:qlssolentrydiff}
\end{lemma}

\begin{proof}
Let us use the same notation as in the proof of Lemma \ref{lem:qlssolentry}. The proof of this lemma is quite similar to that one. The main difference is that here we replace $R$ with a unitary operator $Q$ satisfying 
\bea
&Q \ket{0} \ket{1, -_{i,j}} = \ket{1} \ket{1, -_{i,j}},~~~~~~~~~~~~~~~~~~\label{eq:q1} \\
&Q \ket{0} \ket{1, +_{i,j}} = \ket{0} \ket{1, +_{i,j}}, ~~~~~~~~~~~~~~~~~~\label{eq:q2}\\
&Q \ket{0}  \ket{1, l} = \ket{0} \ket{1, l}, ~1\le l \le N, ~l \neq i, j, \label{eq:q3}\\
&Q \ket{0}  \ket{0, k} = \ket{0} \ket{0, k}, ~1\le k \le N, ~~~~~~~~~
\label{eq:q4}
\eea
where 
\bea
\ket{1, \pm_{i, j}} \defeq \ket{1} \otimes  \dfrac{ \ket{i} \pm \ket{j}}{\sqrt{2}}.
\eea
Then by Eqs.~(\ref{eq:lcu2}), (\ref{eq:q1}), (\ref{eq:q2}), (\ref{eq:q3}) and (\ref{eq:q4}), we get
\bea
QW \ket{0^m}_1 \ket{0}_2  \ket{\bb}_3
&=&\dfrac{\braket{1,-_{i, j}}{\bz'}}{\alpha} \ket{0^m}_1 \ket{1}_2 \ket{1,-_{i,j}}_3 \nonumber \\
&+&\dfrac{\braket{1,+_{i, j}}{\bz'}}{\alpha} \ket{0^m}_1 \ket{0}_2 \ket{1,+_{i,j}}_3 \nonumber \\
&+&\dsum_{l \neq i,j} \dfrac{\braket{1,l}{\bz'}}{\alpha} \ket{0^m}_1 \ket{0}_2 \ket{1,l}_3 \nonumber \\
&+&\dsum_{k} \dfrac{\braket{0,k}{\bz'}}{\alpha} \ket{0^m}_1 \ket{0}_2 \ket{0,k}_3\nonumber \\
&+&\ket{\Xi^{\perp}}_{1,2,3}, 
\eea
where $W$ acts on the first and third registers, $Q$ acts on the second and third registers, and $\ket{\Xi^{\perp}}$ is an unnormalized state satisfying 
$(\ketbra{0^m}{0^m}\otimes I) \ket{\Xi^{\perp}}=0$. If we measure the first $m+1$ qubits of this state, then the probability of getting outcome $0^m1$ is 
\bea
p'' \defeq \dfrac{\abs{\braket{1,-_{i,j}}{\bz'}}^2}{\alpha^2}.
\eea
Recall that $\ket{\bz}=\bA^+ \ket{\bb}=\ket{1}|{\hat{\beta}}\rangle$ and $\ket{\bz'}=h(\bA)\ket{\bb}$ satisfy $\nm{\ket{\bz}-\ket{\bz'}}=\bgo{\epsilon}$. As a result, $\abs{\braket{1,-_{i,j}}{\bz'}}=\alpha \sqrt{p''}$ is an $\bgo{\epsilon}$-additive approximation of $|\braket{1,-_{i,j}}{\bz}|=|{\hat{\beta}_i-\hat{\beta}_j}|/\sqrt{2}$. So in order to estimate $|{\hat{\beta}_i-\hat{\beta}_j}|$ up to additive error $\bgo{\epsilon}$, we only need to obtain an $\bgo{\epsilon}$-additive approximation of $\alpha \sqrt{p''}$. To achieve this, we use amplitude estimation to obtain an $\epsilon''$-additive approximation $\hat{p}''$ of $p''$, where $\epsilon''=\tht{\epsilon^2/\alpha^2}$, succeeding with probability at least $3/4$. Then $\sqrt{\hat{p}''}$ is an $\bgo{\epsilon/\alpha}$-additive approximation of $\sqrt{p''}$, and hence $ \alpha \sqrt{\hat{p}''}$ is an $\bgo{\epsilon}$-additive approximation of $\alpha \sqrt{p''}$, as desired. 

The above basic algorithm has success probability at least $3/4$. To raise the success probability to at least $1-\delta$, we repeat this algorithm $\bgo{\log{1/\delta}}$ times, and take the median of the estimates from these runs. A standard Chernoff's bound ensures that the failure probability is at most $\delta$. 

To analyze the complexity of this algorithm, note that all the parameters are on the same order as in the proof of Lemma \ref{lem:qlssolentry}. Moreover, $Q$ can be implemented in time $\poly{\log{N}}$. Therefore, this algorithm makes 
\bea
\bgo{\dfrac{d^{1.5}\kappa^3}{\epsilon^2} \cdot \poly{\log{\dfrac{\kappa}{\epsilon\delta}} }}
\eea
uses of $\mcal{P}_A$ and $\mcal{P}_b$, and is gate-efficient, as claimed. 
\end{proof} 

Our algorithm for solving the LR-P problem also requires the following procedure for determining whether a given vector $\beta \in \bbR^d$ is close to $\hat{\beta}$ or $-\hat{\beta}$, under the promise that one of these cases holds. 

\begin{lemma}
Let $\bX$, $\by$ and $\hat{\beta}$ be defined as in LR-P. Suppose $\beta \in \bbR^d$ is given such that either $\|{\beta-\hat{\beta}}\| \le \delta$ or $\|{\beta+\hat{\beta}}\| \le \delta$, for some $\delta < \tau/(2\sigma\rho \sqrt{d})$, where $\tau \defeq \tau(\bX,\by)$, $\sigma\defeq \sigma(\bX)$, and $\rho\defeq \rho(\by)$. Then there exits a gate-efficient quantum algorithm that makes $\bgo{d^{1.5}\kappa}$ uses of $\px$ and $\py$, and determines which case holds, succeeding with high probability (e.g. at least $3/4$). 
\label{lem:sgnbeta}
\end{lemma}
\begin{proof}
Let $\hat{\by} \defeq \proj{\bX} \by = \bX \hat{\beta}$. Then since $\tau=\tau(\bX,\by)=\nm{\hat{\by}}^2/\nm{\by}^2=\nm{\hat{\by}}^2$ (recall that $\nm{\by}=1$), we have $\nm{\hat{\by}}=\sqrt{\tau}$. Meanwhile, recall that the singular values of $\bX$ are in the range $[1/\kappa, 1]$. Thus, we have
\bea
\sqrt{\tau} \le \|{\hat{\beta}}\| = \nm{\bX^+ \by} = \nm{\bX^+ \hat{\by}} \le \kappa\sqrt{\tau}.
\label{eq:nmhatbeta}
\eea
Note that $\|{\hat{\beta}- (-\hat{\beta})}\|=2\|{\hat{\beta}}\| \ge 2\sqrt{\tau}$. So by the triangle inequality, at least one of $\|{\beta-\hat{\beta}}\| \ge \sqrt{\tau}$ and $\|{\beta-(-\hat{\beta})}\| \ge \sqrt{\tau}$ must hold. Then since $\delta < \tau/(2\sigma\rho \sqrt{d}) \le \sqrt{\tau}$ (note that $\tau\le 1$ and $\sigma, \rho, d \ge 1$), the two cases $\|{\beta-\hat{\beta}}\| \le \delta$ and $\|{\beta+\hat{\beta}}\| \le \delta$ cannot happen simultaneously. 

Recall that we have shown in the proof of Lemma \ref{lem:hs} that 
\bea
\nm{\bx_i} \le   \dfrac{\sigma \sqrt{d}}{\sqrt{N}}, &~~1 \le i \le N
\label{eq:nmxi}
\eea
(see Eq.~(\ref{eq:nmxibound})). Combining Eqs.~(\ref{eq:nmhatbeta}) and (\ref{eq:nmxi}) yields
\bea
\abs{\bx_i^T \hat{\beta}} \le \dfrac{ \sigma \kappa  \sqrt{\tau d}}{ \sqrt{N}}, &~~1 \le i \le N.
\label{eq:nmxibeta}
\eea
Moreover, by $\nm{\by}=1$ and $\rho(\by)=\rho$, we obtain 
\bea
\abs{y_i} \le \dfrac{\rho}{\sqrt{N}}, &~~1 \le i \le N.
\label{eq:nmyi}
\eea
Now let $\hat{q}_i \defeq y_i \cdot \bx_i^T \hat{\beta}$ for $i \in \intset{N}$. Then Eqs.~(\ref{eq:nmxibeta}) and (\ref{eq:nmyi}) imply that
\bea
\abs{\hat{q}_i} \le \dfrac{\sigma \rho \kappa \sqrt{\tau d}}{N}, & ~~1 \le i \le N,
\label{eq:hatqi}
\eea
Furthermore, we have
\bea
\dsum_{i=1}^N \hat{q}_i
&=& \dsum_{i=1}^N y_i \cdot \bx_i^T \hat{\beta} \\
&=& \by^T \bX \hat{\beta}\\
&=&\by^T \proj{\bX} \by \\
&=& \nm{\hat{\by}}^2 \\
&=&\tau.
\label{eq:sumhatqi}
\eea

Now let $q_i \defeq y_i \cdot \bx_i^T \beta$ for $i \in \intset{N}$. We claim that we can distinguish the cases $\|{\beta-\hat{\beta}}\| \le \delta$ and $\|{\beta+\hat{\beta}}\| \le \delta$ by estimating the quantity $\sum_{i=1}^N q_i$ up to additive error $\tau/2$. To prove this, let us consider these two cases separately:
\bit
\item Case 1: $\|{\beta-\hat{\beta}}\| \le \delta < \tau/(2\sigma \rho \sqrt{d})$. Using Eqs.~(\ref{eq:nmxi}) and (\ref{eq:nmyi}), we get
\bea
\abs{q_i-\hat{q}_i} 
&=&\abs{y_i \cdot \bx_i^T\lr{\beta - \hat{\beta}}}\\
&\le & \abs{y_i} \nm{\bx_i} \|{\beta-\hat{\beta}}\| \\
&\le & \dfrac{\rho}{\sqrt{N}} \cdot \dfrac{\sigma\sqrt{d}}{\sqrt{N}} \cdot \delta\\
&<&  \dfrac{\tau}{2N},
\label{eq:qihatqipos}
\eea
for all $i \in \intset{N}$. Then by Eqs.~(\ref{eq:hatqi}) and (\ref{eq:qihatqipos}), we find that
\bea
\abs{q_i} &<& \abs{\hat{q}_i} + \abs{q_i - \hat{q}_i} \\
&\le & \dfrac{\sigma \rho \kappa \sqrt{\tau d}} {N} + \dfrac{\tau}{2N} \\
&\le &\dfrac{2\sigma \rho \kappa \sqrt{d}} {N},
\label{eq:absqipos}
\eea
for all $i \in \intset{N}$ (note that $\rho, \sigma, \kappa, d \ge 1$ and $\tau \le 1$). Furthermore, Eqs.~(\ref{eq:sumhatqi}) and (\ref{eq:qihatqipos}) imply that
\bea
\dsum_{i=1}^N q_i &\ge & \dsum_{i=1}^N \hat{q}_i - \dsum_{i=1}^N \abs{q_i-\hat{q}_i}  \\
&>& 
 \tau -  \dfrac{\tau}{2} \\ 
& =& \dfrac{\tau}{2}.
\label{eq:sumqipos}
\eea

\item Case 2: $\|{\beta+\hat{\beta}}\| \le \delta < \tau/(2\sigma \rho \sqrt{d})$. Using Eqs.~(\ref{eq:nmxi}) and (\ref{eq:nmyi}), we get
\bea
\abs{q_i+\hat{q}_i} 
&=&\abs{y_i \cdot \bx_i^T\lr{\beta + \hat{\beta}}}\\
&\le & \abs{y_i} \nm{\bx_i} \|{\beta + \hat{\beta}}\| \\
&\le & \dfrac{\rho}{\sqrt{N}} \cdot \dfrac{\sigma\sqrt{d}}{\sqrt{N}} \cdot \delta\\
&<&  \dfrac{\tau}{2N},
\label{eq:qihatqineg}
\eea
for all $i \in \intset{N}$. Then by Eqs.~(\ref{eq:hatqi}) and (\ref{eq:qihatqineg}), we find that
\bea
\abs{q_i} &<& \abs{\hat{q}_i} + \abs{q_i + \hat{q}_i} \\
&\le & \dfrac{\sigma \rho \kappa \sqrt{\tau d}} {N} + \dfrac{\tau}{2N} \\
&\le &\dfrac{2\sigma \rho \kappa \sqrt{d}} {N},
\label{eq:absqineg}
\eea
for all $i \in \intset{N}$ (note that $\rho, \sigma, \kappa, d \ge 1$ and $\tau \le 1$). Furthermore, Eqs.~(\ref{eq:sumhatqi}) and (\ref{eq:qihatqineg}) imply that
\bea
\dsum_{i=1}^N q_i &\le & -\dsum_{i=1}^N \hat{q}_i + \dsum_{i=1}^N \abs{q_i+\hat{q}_i}  \\
&<&  -\tau + \dfrac{\tau}{2} \\ 
& =& -\dfrac{\tau}{2}.
\label{eq:sumqineg}
\eea
\eit
Comparing Eqs.~(\ref{eq:sumqipos}) and (\ref{eq:sumqineg}), we know that we can distinguish the two cases $\|{\beta-\hat{\beta}}\| \le \delta$ and $\|{\beta+\hat{\beta}}\| \le \delta$ by estimating $\sum_{i=1}^N q_i$ up to additive error $\tau/2$, as claimed. 

We obtain a $\tau/2$-additive approximation of $\sum_{i=1}^N q_i$ as follows. Let $U$ be a unitary operator such that
\bea
U \ket{i} \ket{0}
= \ket{i} \ket{\psi_i}, & ~~~1 \le i \le N,
\eea
where  
\bea
\ket{\psi_i} \defeq { \sqrt{\dfrac{1}{2}+ \dfrac{N q_i}{2\Delta}} \ket{0}
+\sqrt{\dfrac{1}{2}- \dfrac{N q_i}{2\Delta}} \ket{1}}
\eea
in which $\Delta \defeq 2 \sigma \rho \kappa \sqrt{d}$. Note that $U$ is a valid unitary operator, since $N \abs{q_i} \le \Delta$ by Eqs.~(\ref{eq:absqipos}) and (\ref{eq:absqineg}). Then we have
\bea
U \lb \dfrac{1}{\sqrt{N}} \dsum_{i=1}^N \ket{i} \rb \ket{0}
= \dfrac{1}{\sqrt{N}} \dsum_{i=1}^N \ket{i} \ket{\psi_i}.
\eea
If we measure the second register of this state in the standard basis, then the probability of obtaining outcome $0$ is 
\bea
p \defeq \dfrac{1}{2}+\dfrac{\sum_{i=1}^N q_i}{\Delta}.
\eea
We use amplitude estimation to obtain an $\tau/(2\Delta)$-additive approximation $\hat{p}$ of $p$, succeeding with high probability (e.g. at least $3/4$). Then $ (\hat{p}-1/2) \Delta $ is a $\tau/2$-additive approximation of $\sum_{i=1}^N q_i$, as desired. 

The unitary operator $U$ can be implemented as follows. For any $i \in \intset{N}$,
given the state $\ket{i}\ket{0}$, we first transform it into $\ket{i} \ket{0} \ket{q_i}$, where $q_i=y_i (\sum_{j=1}^d x_{i,j} \beta_j)$ can be computed by making $\bgo{d}$ uses of $\px$ and $\py$. Then we perform the controlled-rotation
\bea
\ket{0} \ket{q_i}
\to 
\ket{\psi_i} \ket{q_i}
\eea
on the last two registers. After that, we uncompute $q_i$ in the last register by making $\bgo{d}$ uses of $\px$ and $\py$, and get the desired state $\ket{i} \ket{\psi_i}$. This implemenation of $U$ requires $\bgo{d}$ uses of $\px$ and $\py$, and is gate-efficient. 

Since we want to estimate $p$ up to additive error $\tau/(2\Delta)$, amplitude estimation requires 
\bea
\bgo{\dfrac{\Delta}{\tau}}=\bgo{\dfrac{\sigma \rho \kappa \sqrt{d}}{\tau}}=\bgo{\kappa \sqrt{d}}
\eea
repetitions of $U$ (recall that $\sigma=\bgo{1}$, $\rho=\bgo{1}$ and $\tau=\omg{1}$). As a result, this algorithm makes $\bgo{d^{1.5}\kappa }$ uses of $\px$ and $\py$, and is gate-efficient, as claimed.

\end{proof}

Now we are ready to state our algorithm for solving the LR-P problem.

\begin{theorem}
The LR-P problem can be solved by a gate-efficient quantum algorithm that makes 
$$\bgo{ \dfrac{d^{2.5} \kappa^3}{\delta^2}   
 \cdot \poly{\log{\dfrac{d \kappa}{\delta}}}}$$ 
uses of $\px$ and $\py$, where $\delta \defeq \mmin{\epsilon, 1/d}$.
\label{thm:lrfit}
\end{theorem}

\begin{proof}
\tbf{Algorithm:}
Let $\bX$ and $\by$ be defined as in LR-P. Let $\tau \defeq \tau(\bX,\by)=\omg{1}$, $\sigma\defeq \sigma(\bX)=\bgo{1}$ and $\rho\defeq \rho(\by)=\bgo{1}$. We use the following algorithm to obtain a vector $\beta=(\beta_1, \beta_2, \dots, \beta_d)^T \in \bbR^d$ satisfying $\infnm{\beta-\hat{\beta}} \le \epsilon$, succeeding with probability at least $2/3$:
\ben
\item Let $\epsilon' \defeq \mmin{\tau/(2\sigma \rho d), \epsilon}$.

\item For each $j \in \intset{d}$, we run the algorithm in Lemma  \ref{lem:qlssolentry} to obtain an $\epsilon'/6$-additive approximation $\mu_j$ of $|{\hat{\beta}_j}|$, succeeding with probability at least $1-1/(25d)$. 

\item Let $S \defeq \seta{j \in \intset{d}  :~\mu_j>2\epsilon'/3}$. If $S=\emptyset$, then this algorithm fails; otherwise, we continue as follows.

\item Pick arbitrary $j_0 \in S$. For each $j \in S$, $j \neq j_0$, we run the algorithm in Lemma \ref{lem:qlssolentrydiff} to obtain an $\epsilon'/6$-additive approximation $\gamma_j$ of $|\hat{\beta}_{j_0} - \hat{\beta}_j |$, succeeding with probability at least $1-1/(25d)$. 

\item For each $j \in \intset{d}$, we define $s_j \in \seta{-1, 0, 1}$ as follows:
\bit
\item If $j \not\in S$, then $s_j=0$.
\item If $j=j_0 \in S$, then $s_j=1$. 
\item Otherwise, we have $j \in S$ and $j \neq j_0$. If $\abs{|\mu_{j_0}-\mu_j|-\gamma_j} \le \epsilon'/2$, then $s_j=1$; otherwise,  $s_j=-1$.
\eit

\item Let $\beta' = (\beta_1',\beta_2',\dots,\beta_d')^T \in \bbR^d$ be defined as $\beta_j' \defeq s_j \mu_j$ for each $ j \in \intset{d}$. We will prove below that, with high probability, either $\|\beta' - \hat{\beta}\| < \tau/(2\sigma\rho\sqrt{d})$ or $\|\beta' + \hat{\beta}\| < \tau/(2\sigma\rho\sqrt{d})$.
We run the algorithm in Lemma \ref{lem:sgnbeta} to determine which case holds, succeeding with probability at least $3/4$. If the first case holds, then we return $\beta \defeq \beta'$ as our estimate of $\hat{\beta}$; otherwise, we return $\beta \defeq -\beta'$ as our estimate of $\hat{\beta}$.\\
\een

\tbf{Correctness:} Let us call the case where all the instances of the algorithms in Lemmas \ref{lem:qlssolentry}, \ref{lem:qlssolentrydiff} and \ref{lem:sgnbeta} succeed the \emph{typical} case. By union bound, the probability of this case   happening is at least $1-2d/(25d)-1/4>2/3$. We will prove that in the typical case, our algorithm outputs a correct $\beta$ (i.e. $\infnm{\beta-\hat{\beta}} \le \epsilon$) with certainty. 

In the typical case, we have 
\bea
\abs{\mu_i - \abs{\hat{\beta}_i}} \le \dfrac{\epsilon'}{6}, & ~~1\le i \le d,
\label{eq:mui}
\eea
and
\bea
~~~~~\abs{\gamma_j - \abs{\hat{\beta}_{j_0}-\hat{\beta}_j }} \le \dfrac{\epsilon'}{6}, & ~~\forall j \in S, j \neq j_0. ~
\label{eq:zetaj}
\eea
Then using the definition of $S$, we get
\bea
\abs{\hat{\beta}_j} 
&\ge &  \abs{\mu_j} - \abs{\mu_i - \abs{\hat{\beta}_i}} \\
&>& \dfrac{2\epsilon'}{3}-\dfrac{\epsilon'}{6} \\
&=& \dfrac{\epsilon'}{2},  ~~~\forall j \in S, \label{eq:betajnorm1}
\eea
and
\bea
\abs{\hat{\beta}_j} 
&\le &  \abs{\mu_j} + \abs{\mu_i - \abs{\hat{\beta}_i}} \\
&\le & \dfrac{2\epsilon'}{3}+\dfrac{\epsilon'}{6}\\
&=& \dfrac{5\epsilon'}{6},  ~~~\forall j \not\in S. \label{eq:betajnorm2}
\eea

Recall that we have shown in the proof of Lemma \ref{lem:sgnbeta} that 
\bea
\nm{\hat{\beta}} = \sqrt{\dsum_{i=1}^d |\hat{\beta}_i|^2 }\ge \sqrt{\tau}
\eea
(see Eq.~(\ref{eq:nmhatbeta})). This implies that there exists some $i_0 \in \intset{d}$ such that 
\bea
\abs{\hat{\beta}_{i_0}} \ge \sqrt{\dfrac{\tau}{d}} \ge \dfrac{\tau}{\sigma \rho d}
\ge 2\epsilon'
\label{eq:hatbetai0}
\eea
(note that $\sigma, \rho, d \ge 1$ and $\tau \le 1$). Then by Eqs.~(\ref{eq:mui}) and (\ref{eq:hatbetai0}), we obtain
\bea
\mu_{i_0} &\ge & \abs{\beta_{i_0}} - \abs{\mu_{i_0} - \abs{\hat{\beta}_{i_0}}} \\
&\ge & 2\epsilon'- \dfrac{\epsilon'}{6} \\
&>& \dfrac{2\epsilon'}{3}.
\eea
Thus, we have $i_0 \in S$ and $S \neq \emptyset$. So our algorithm does not fail in the typical case.

Now we claim that $s_j=\mrm{sgn}(\hat{\beta}_j)\cdot \mrm{sgn}(\hat{\beta}_{j_0})$ for any $j \in S$. The proof is as follows.
\bit
\item If $j=j_0$, then $s_{j}=1$ by definition.
\item If $j\neq j_0$ and $\mrm{sgn}(\hat{\beta}_j)=\mrm{sgn}(\hat{\beta}_{j_0})$, then we have
\bea
\abs{\hat{\beta}_{j_0}-\hat{\beta}_j}=\abs{\abs{\hat{\beta}_{j_0}}-\abs{\hat{\beta}_j}}.
\label{eq:hatbetaj0j1}
\eea
Combining Eqs.~(\ref{eq:mui}), (\ref{eq:zetaj}) and (\ref{eq:hatbetaj0j1}) gives
\bea
\abs{\abs{\mu_{j_0}-\mu_j} - \gamma_j}
&\le & 
\abs{\mu_{j_0}-\abs{\hat{\beta}_{j_0}}} \nonumber \\
&+&
\abs{\mu_{j}-\abs{\hat{\beta}_j}} \nonumber \\
&+&
\abs{\gamma_{j}-\abs{\hat{\beta}_{j_0} - \hat{\beta}_{j}}}
\\
&\le & \dfrac{ \epsilon'}{6}+\dfrac{ \epsilon'}{6}+\dfrac{ \epsilon'}{6}\\
&=&\dfrac{\epsilon'}{2}. 
\eea
This implies that $s_j=1$ for this $j$.

\item If $j\neq j_0$ and $\mrm{sgn}(\hat{\beta}_j) = -\mrm{sgn}(\hat{\beta}_{j_0})$, then we have
\bea
\abs{\hat{\beta}_{j_0}-\hat{\beta}_j}
&=&\abs{\hat{\beta}_{j_0}}+\abs{\hat{\beta}_j} \nonumber \\
&= &
\abs{\abs{\hat{\beta}_{j_0}}-\abs{\hat{\beta}_j}} \\
&+& 2 \mmin{\abs{\hat{\beta}_{j_0}},\abs{\hat{\beta}_j}} \nonumber \\
& > &
\abs{\abs{\hat{\beta}_{j_0}}-\abs{\hat{\beta}_j}} + \epsilon',
\label{eq:hatbetaj0j2}
\eea
since $|\hat{\beta}_{j_0}|,|\hat{\beta}_j| > \epsilon'/2$ by Eq.~(\ref{eq:betajnorm1}). Combining Eqs.~(\ref{eq:mui}), (\ref{eq:zetaj}) and  (\ref{eq:hatbetaj0j2}) yields
\bea
\abs{\abs{\mu_{j_0}-\mu_j} - \gamma_j} &\ge &
\epsilon'
-\abs{\mu_{j_0}-\abs{\hat{\beta}_{j_0}}} \nonumber \\
&-&
\abs{\mu_{j}-\abs{\hat{\beta}_j}} \nonumber \\
&-&
\abs{\gamma_{j}-\abs{\hat{\beta}_{j_0} - \hat{\beta}_{j}}}
\\
&>& \epsilon' - \dfrac{ \epsilon'}{6} -\dfrac{ \epsilon'}{6}-\dfrac{ \epsilon'}{6}\\
&=&\dfrac{\epsilon'}{2}.
\eea
This implies that $s_j=-1$ for this $j$.
\eit

The fact that $s_j=\mrm{sgn}(\hat{\beta}_j)\cdot \mrm{sgn}(\hat{\beta}_{j_0})$
for all $j \in S$ implies that either 
\bea
\mrm{sgn}(\beta_j')=\mrm{sgn}(\hat{\beta}_j), ~~~\forall j \in S,
\eea
or 
\bea
\mrm{sgn}(\beta_j')=-\mrm{sgn}(\hat{\beta}_j), ~~~ \forall j \in S.
\eea
Moreover, by Eq.~(\ref{eq:mui}), we know that 
\bea
\abs{|{\beta_j'}|-|{\hat{\beta}_j}|} \le \dfrac{\epsilon'}{6},~~~\forall j \in S.
\eea 
As a result, we have either 
\bea
\abs{{\beta_j' - \hat{\beta}_j}} \le  \dfrac{ \epsilon'}{6}, ~~~\forall j \in S,
\eea
or 
\bea
\abs{{\beta_j' +\hat{\beta}_j}} \le \dfrac{ \epsilon'}{6}, ~~~\forall j \in S.
\eea

Meanwhile, for any $j \not\in S$, we have $s_j=0$ and $|{\hat{\beta}_j}| \le {5\epsilon'}/{6}$ by Eq.~(\ref{eq:betajnorm2}). It follows that $\beta_j'=0$ and \bea
\abs{{\beta_j' -\hat{\beta}_j}}= \abs{{\beta_j' + \hat{\beta}_j}}
\le \dfrac{5\epsilon'}{6},~~\forall j \not\in S.
\eea

Combining the cases $j \in S$ and $j \not\in S$, we know that either 
\bea
\infnm{\beta' - \hat{\beta}} \le \dfrac{5\epsilon'}{6}<\epsilon'
\eea
or 
\bea
\infnm{\beta' +\hat{\beta}} \le \dfrac{5\epsilon'}{6}<\epsilon'.
\eea 
As a result, we have either 
\bea
\|{\beta' - \hat{\beta}}\| < \sqrt{d}\epsilon'\le \dfrac{\tau}{2\sigma \rho \sqrt{d}}
\eea 
or 
\bea
\|{\beta' +\hat{\beta}}\| < \sqrt{d}\epsilon' \le \dfrac{\tau}{2\sigma \rho \sqrt{d}}.
\eea
In the typical case, our algorithm in Lemma \ref{lem:sgnbeta} correctly determines which case holds. If the first case holds, then it outputs $\beta=\beta'$ which satisfies $\infnm{\beta - \hat{\beta}} <\epsilon' \le \epsilon$; otherwise, it outputs $\beta=-\beta'$ which also satisfies $\infnm{\beta - \hat{\beta}} <\epsilon' \le \epsilon$, as desired.\\

\tbf{Complexity:}  Recall that $\epsilon'=\mmin{\tau/(2\sigma\rho d), \epsilon}$ and $\delta=\mmin{1/d, \epsilon}$, where $\tau=\omg{1}$, $\sigma=\bgo{1}$ and
$\rho=\bgo{1}$. So we have $\epsilon'=\omg{\delta}$. Let us analyze the complexity of each step. Step 2 makes $\bgo{d}$ uses of the algorithm in Lemma \ref{lem:qlssolentry}, so it requires 
\bea
&&\bgo{d \cdot \dfrac{d^{1.5}\kappa^3}{(\epsilon')^2} \cdot \poly{\log{\dfrac{d\kappa}{\epsilon'}}}} \\
&=&
\bgo{\dfrac{d^{2.5}\kappa^3}{\delta^2} \cdot \poly{\log{\dfrac{d\kappa}{\delta}}}}
\eea
uses of $\px$ and $\py$, and is gate-efficient. Step 4 makes $\bgo{d}$ uses of the algorithm in Lemma \ref{lem:qlssolentrydiff}, so it requires 
\bea
&&\bgo{d \cdot \dfrac{d^{1.5}\kappa^3}{(\epsilon')^2} \cdot \poly{\log{\dfrac{d\kappa}{\epsilon'}}}}\\
&=&
\bgo{\dfrac{d^{2.5}\kappa^3}{\delta^2} \cdot \poly{\log{\dfrac{d\kappa}{\delta}}}}
\eea
uses of $\px$ and $\py$, and is gate-efficient. Step 6 makes $\bgo{1}$ uses of the algorithm in Lemma \ref{lem:sgnbeta}, so it requires
$\bgo{\kappa d^{1.5}}$ uses of of $\px$ and $\py$, and is gate-efficient. 
Furthermore, the classical computation in this algorithm takes $\bgo{d}$ time. As a result, this algorithm makes 
\bea
\bgo{\dfrac{d^{2.5}\kappa^3}{\delta^2} \cdot \poly{\log{\dfrac{d\kappa}{\delta}}}}
\eea
uses of $\px$ and $\py$, and is gate-efficient, as claimed.
\end{proof}

Our algorithm for computing $\hat{\beta}=\bX^+ \by$ is more efficient than an alternative one in which one first creates multiple copies of the state proportional to $\hat{\beta}$ and then uses statistical sampling and quantum state tomography to determine the $\hat{\beta}_j$'s (as suggested by Ref.~\cite{wiebe2012quantum}). The main reason is that, in order to obtain an $\epsilon$-additive approximation of $|\hat{\beta}_j|^2$, the sampling-based approach would require $\bgo{1/\epsilon^2}$ copies of the state encoding $\hat{\beta}$, but amplitude estimation only needs $\bgo{1/\epsilon}$ repetitions of the procedure for preparing this state. So it is more efficient to couple the state generation process with amplitude estimation (as we did in our algorithm) rather than statistical sampling. 

We also remark that the algorithm in Lemma \ref{lem:qlssolentry} can be modified to produce a quantum state approximately proportional to $\hat{\beta}$. Specifically, note that if we measure the first register of $W\ket{0^m}\ket{\bb}$ (in Eq.~(\ref{eq:lcu2})) in the standard basis, then conditioning on the outcome being $0^m$, we would obtain the normalized version of $h(\bA)\ket{\bb}$, which is close to the normalized version of $\bA^+\ket{\bb}=\ket{1}|{\hat{\beta}}\rangle$. The probablity of this event happening is $\nm{h(\bA)\ket{\bb}}^2/\alpha^2=\omg{1/\alpha^2}$. We can use amplitude amplification to raise this probability to $\omg{1}$, which requires $\bgo{\alpha}$ repetitions of $W$ and the procedure for preparing $\ket{\bb}$. This leads to a gate-efficient algorithm that makes 
\bea
\bgo{d^{1.5}\kappa^2 \cdot \poly{\log{\dfrac{\kappa}{\epsilon}} }}
\eea
uses of $\px$ and $\py$, and prepares a quantum state $\epsilon$-close to $\frac{|\hat{\beta}\rangle}{\nm{|\hat{\beta}\rangle}}$ in $l^2$ norm, succeeding with probability $\omg{1}$ (with a flag indicating success). By utilizing Ambainis'  \emph{variable-time amplitude amplification} \cite{ambainis2012variable}, we can reduce the $\kappa$-dependence from quadratic to linear, as done in Section 5 of Ref.~\cite{childs2015quantum}. This leads to a gate-efficient algorithm with query complexity 
\bea
\bgo{d^{1.5}\kappa \cdot \poly{\log{\dfrac{\kappa}{\epsilon}} }}
\eea
for the same task.

One may compare this algorithm for preparing a quantum state approximately proportional to the optimal parameters 
\beq
\hat{\beta}=\bX^+ \by=(\bX^T \bX)^{-1}\bX^T \by
\eeq
with the one in Ref.~\cite{wiebe2012quantum} for the same task. Our algorithm is based on the singular value decomposition (SVD) of $\bX$, and it applies $\bX^+$ to $\by$ in a direct manner. Consequently, it has only linear dependence on the condition number $\kappa$ of $\bX$. By contrast, Ref.~\cite{wiebe2012quantum} needs to first apply $\bX^T$ to $\by$, which incurs a $\kappa$ factor in the complexity; then it needs to apply $(\bX^T \bX)^{-1}$ to the output of the first step, which incurs another $\kappa^2$ factor in the complexity. So its overall complexity is at least cubic in $\kappa$. This means that our algorithm has polynomially better dependence on $\kappa$ than the one in Ref.~\cite{wiebe2012quantum}. Furthermore, due to the fact we use the new strategy of Ref.~\cite{childs2015quantum} for matrix inversion, our algorithm also has exponential better dependence on the desired precision $\epsilon$ in the output state.

\section{Estimating the Quality of the Least-Squares Fit}
\label{sec:lrtest}
In this section, we describe a quantum algorithm for solving the LR-Q problem, i.e. estimating the quality $\tau=\|{\bX \hat{\beta}}\|^2/\nm{\by}^2$ of the least-squares fit $\by \approx \bX \hat{\beta}$ for a given data set $(\bX, \by)$ (without computing the parameters $\hat{\beta}$ explicitly). This algorithm requires the following variant of phase estimation  \cite{kitaev1995quantum,cleve1998quantum}, which decides whether the eigenphase corresponding to an eigenvector of a unitary operator is $\theta$ or far away from $\theta$, for some given $\theta \in [0, 2\pi)$, succeeding with probability close to $1$. (Similar procedures have been used in Refs.~\cite{nagaj2009fast,wang2013efficient,childs2015quantum}.)

\begin{lemma}
Let $U$ be a unitary operator with eigenvectors $\ket{\psi_j}$ such that 
$U \ket{\psi_j} = e^{i \theta_j} \ket{\psi_j}$ for some $\theta_j \in [0, 2\pi)$. Let $\theta \in [0, 2\pi)$ and let $\Delta$, $\delta \in (0, 1)$. Then there is a unitary procedure $\mcal{P}$ that makes $\bgo{(1/\Delta)\cdot \log{1/\delta}}$ uses of $U$, and uses $\poly{\log{1/(\Delta\delta)}}$ additional 2-qubit gates, and satisfies
\bea
\mcal{P} \ket{0}\ket{0^l} \ket{\psi_j} = \lb \alpha_{j,0} \ket{0} \ket{\eta_{j, 0}} + \alpha_{j, 1} \ket{1} \ket{\eta_{j, 1}} \rb \ket{\psi_j},
\label{eq:p00lpsi}
\eea
where $l=\bgo{\log{1/\Delta} \log{1/\delta}}$, $\abs{\alpha_{j, 0}}^2 + \abs{\alpha_{j, 1}}^2=1$, $\ket{\eta_{j, 0}}$ and $\ket{\eta_{j, 1}}$ are two normalized states, and
\bit
\item If $\theta_j=\theta$, then $\abs{\alpha_{j, 0}}^2 \ge 1-\delta$.
\item If $\abs{\theta_j-\theta} \ge \Delta$, then $\abs{\alpha_{j, 1}}^2 \ge 1-\delta$.
\eit
\label{lem:boostedpe}
\end{lemma}
\begin{proof}
We can get a $\Delta/2$-additive approximation of $\theta_j$ by using the standard phase estimation, which makes $\bgo{1/\Delta}$ uses of $U$ and uses $\poly{\log{1/\Delta}}$ additional 2-qubit gates. This is sufficient to distinguish  the two cases. However, it only succeeds with probability $\omg{1}$. To raise this probability to at least $1-\delta$, we repeat this procedure $\bgo{\log{1/\delta}}$ times and check whether the median of the estimates is $\Delta/2$-close to $\theta$. A standard Chernoff's bound ensures that the failure proability is at most $\delta$. This boosted procedure, denoted by $\mcal{P}$, makes $\bgo{(1/\Delta) \cdot \log{1/\delta}}$ uses of $U$, and uses $\poly{\log{1/(\Delta\delta)}}$ additional 2-qubit gates, and satisfies all the desired properties. 
\end{proof}

\begin{theorem}
The LR-Q problem can be solved by a gate-efficient quantum algorithm that makes 
$$\bgo{\dfrac{d^{1.5} \kappa }{\epsilon} \cdot \poly{\log{\dfrac{ \kappa }{\epsilon}}}}$$
uses of $\px$ and $\py$.
\label{thm:lrtest}
\end{theorem}

\begin{proof}
\noindent\tbf{Algorithm:}
Let $\bX$ and $\by$ be defined as in LR-Q. We use the following algorithm to obtain an ${\epsilon}$-additive approximation of $\tau=\nm{\proj{\bX}\by}^2/\nm{\by}^2=\nm{\proj{\bX} \by}^2$ (recall that $\nm{\by}=1$), succeeding with probability at least $2/3$. Let $\bA \defeq  \ketbra{1}{0} \otimes \bX^T + \ketbra{0}{1} \otimes \bX$ and $\ket{\bb} \defeq \ket{0}\ket{\by}$. Let $\mcal{P}$ be the unitary procedure in Lemma \ref{lem:boostedpe} for $U=e^{-i\bA}$, $\theta=0$, $\Delta=1/(2\kappa)$ and $\delta={\epsilon/2}$. Suppose
\beq
\mcal{P} \ket{0}_1 \ket{0^l}_2 \ket{\bb}_3 = 
\mu_0 \ket{0}_1 \ket{\vphi_0}_{2,3}
+
\mu_1 \ket{1}_1 \ket{\vphi_1}_{2,3},~
\eeq
where $l=\bgo{\log{1/\Delta}  \log{1/\delta}}$, $\abs{\mu_0}^2+\abs{\mu_1}^2=1$, and $\ket{\vphi_0}_{2,3}$ and $\ket{\vphi_1}_{2,3}$ are some normalized states on the second and third registers. We use amplitude estimation to get an ${\epsilon/2}$-additive approximation $\hat{r}$ of $r\defeq \abs{\mu_1}^2$, succeeding with probability at least $3/4$. Then we return $\hat{r}$ as our estimate of $\tau$. During this process, we use the procedure in Lemma \ref{lem:hs} to implement
$U=e^{-i\bA}$ with precision $\bgo{\epsilon^2/\kappa^2}$ (and failure probability $\bgo{\epsilon^2/\kappa^2}$), and use the procedure in Lemma \ref{lem:preparey} to prepare $\ket{\by}$ with precision $\bgo{\epsilon^2}$.\\

\noindent\tbf{Correctness:} Suppose $\bX$ has the singular value decomposition
\bea
\bX = \dsum_{j=1}^d s_j \ketbra{\bu_j}{\bv_j},
\eea
where $s_j \in [1/\kappa, 1]$, $\ket{\bu_j} \in \bbR^{N}$ and $\ket{\bv_j} \in \bbR^d$ are unit vectors, for all $j \in \intset{d}$. Then $\bA$ has the spectral decomposition
\bea
\bA = \dsum_{j=1}^d s_j  \ketbra{+_j}{+_j} - \dsum_{j=1}^d s_j  \ketbra{-_j}{-_j} ,
\eea
where 
\bea
\ket{\pm_{j}} \defeq \dfrac{1}{\sqrt{2}} \lb \ket{0}\ket{\bu_j} \pm  \ket{1} \ket{\bv_j} \rb.
\label{eq:ketpmj}
\eea
Meanwhile, we can write $\ket{\by}$ as
\bea
\ket{\by} = \dsum_{j=1}^d \alpha_j \ket{\bu_j} + \alpha \ket{\Phi^{\perp}},
\label{eq:ketbydecomp}
\eea
where $\sum_{j=1}^d \abs{\alpha_j}^2 + \abs{\alpha}^2=1$, and $\ket{\Phi^{\perp}}$ is some normalized state satisfying $\braket{\bu_j}{\Phi^{\perp}} = 0$ for all $j$. Note that
\bea
\tau = \nm{\proj{\bX} \by}^2 = \dsum_{j=1}^d \abs{\alpha_j}^2.
\label{eq:taualpha}
\eea
By Eqs.~(\ref{eq:ketpmj}) and (\ref{eq:ketbydecomp}), we obtain 
\bea
\ket{\bb}
&=& \ket{0}\ket{\by}\\
&=& \dsum_{j=1}^d \alpha_j \ket{0}\ket{\bu_j} + \alpha\ket{0} \ket{\Phi^{\perp}} \\
&=& \dsum_{j=1}^d \dfrac{\alpha_j}{\sqrt{2}} \lb \ket{+_j} + \ket{-_j} \rb  + \alpha\ket{0} \ket{\Phi^{\perp}}.
\eea
Note that $\ket{0}\ket{\Phi^{\perp}}$ is an eigenvector of $\bA$ with eigenvalue $0$, i.e. $\bA \ket{0}\ket{\Phi^{\perp}} = 0$. 

Now, since the eigenphase gap around $0$ of $U=e^{-i\bA}$ is at least $1/\kappa$, by Lemma \ref{lem:boostedpe} and our choice of parameters, we get
\beq
\mcal{P}\ket{0}\ket{0^l}\ket{+_j}
= \lb \gamma_{j, 0}^{+}\ket{0} \ket{\phi_{j,0}^{+}}+\gamma_{j, 1}^{+}\ket{1} \ket{\phi_{j,1}^{+}} \rb \ket{+_j},~
\eeq
\beq
\mcal{P}\ket{0}\ket{0^l}\ket{-_j}
= \lb \gamma_{j, 0}^{-}\ket{0} \ket{\phi_{j,0}^{-}}+\gamma_{j, 1}^{-}\ket{1} \ket{\phi_{j,1}^{-}} \rb \ket{-_j},~
\eeq
where $\abs{\gamma_{j,1}^{\pm}}^2 \ge 1-\delta$,
$\abs{\gamma_{j,0}^{\pm}}^2 \le \delta$, $\ket{\phi_{j,0}^{\pm}}$ and
$\ket{\phi_{j,1}^{\pm}}$ are some normalized states, for all $j \in \intset{d}$, and
\beq
\mcal{P}\ket{0}\ket{0^l}\ket{0}\ket{\Phi^{\perp}}
=\lb \eta_0 \ket{0} \ket{\psi_0} +\eta_1 \ket{1} \ket{\psi_1} \rb \ket{0} \ket{\Phi^{\perp}},~
\eeq
where $\abs{\eta_0}^2 \ge 1-\delta$, $\abs{\eta_1}^2 \le \delta$,
$\ket{\psi_{0}}$ and $\ket{\psi_{1}}$ are some normalized states. As a result, we have
\bea
\mcal{P}\ket{0}\ket{0^l}\ket{\bb}
&=&
\dsum_{j=1}^d \dfrac{\alpha_j}{\sqrt{2}} \lb \gamma_{j, 0}^{+}\ket{0} \ket{\phi_{j,0}^{+}}+\gamma_{j, 1}^{+}\ket{1} \ket{\phi_{j,1}^{+}} \rb \ket{+_j} \nonumber\\
&+&
\dsum_{j=1}^d \dfrac{\alpha_j}{\sqrt{2}} \lb \gamma_{j, 0}^{-}\ket{0} \ket{\phi_{j,0}^{-}}+\gamma_{j, 1}^{-}\ket{1} \ket{\phi_{j,1}^{-}} \rb \ket{-_j}\nonumber\\
&+&
\alpha \lb \eta_0 \ket{0} \ket{\psi_0} +\eta_1 \ket{1} \ket{\psi_1} \rb \ket{0} \ket{\Phi^{\perp}}.
\eea
It follows that
\bea
r=\dfrac{1}{2}\dsum_{j=1}^{d} \abs{\alpha_j}^2 \lb 
\abs{ \gamma_{j,1}^{+}}^2 +\abs{ \gamma_{j,1}^{-}}^2 \rb
+ \abs{\alpha}^2 \abs{\eta_{1}}^2.~~~~~~~
\label{eq:ralpha}
\eea
Note that since $|\gamma_{j,1}^{\pm}|^2 \approx 1$ and $\abs{\eta_1}\approx 0$, we have $r \approx \tau$ by Eqs.~(\ref{eq:taualpha}) and (\ref{eq:ralpha}). More precisely, the difference between $r$ and $\tau$ can be bounded using the triangle inequality:
\bea
\abs{r -\tau}
&\le & 
\dfrac{1}{2}\dsum_{j=1}^{d} \abs{\alpha_j}^2 
\lb 1-\abs{ \gamma_{j,1}^{+}}^2 \rb \nonumber\\
&+&\dfrac{1}{2}\dsum_{j=1}^{d} \abs{\alpha_j}^2 
\lb 1- \abs{ \gamma_{j,1}^{-}}^2 \rb \nonumber\\
&+& \abs{\alpha}^2 \abs{\eta_{1}}^2 \\
& \le  &
\dfrac{1}{2}\dsum_{j=1}^{d} \abs{\alpha_j}^2 \cdot \delta
+\dfrac{1}{2}\dsum_{j=1}^{d} \abs{\alpha_j}^2 \cdot \delta \nonumber\\
&+& \abs{\alpha}^2 \cdot \delta \\
& = & \delta \\
& = & \dfrac{\epsilon}{2}.
\eea
Namely, $r$ is an $\epsilon/2$-additive approximation of $\tau$. Meanwhile, $\hat{r}$ is an ${\epsilon/2}$-additive approximation of $r$. It follows that $\hat{r}$ is an ${\epsilon}$-additive approximation of $\tau$, as desired. 

In the above argument, we have ignored the error in the implementation of $U=e^{-i\bA}$ and the error in the preparation of $\ket{\by}$. We will show below that our algorithm only makes $\lto{\kappa^2/\epsilon^2}$ uses of $U$ and $\lto{1/\epsilon^2}$ uses of the procedure for preparing $\ket{\by}$. Thus, provided that $U$ is implemented with precision $\bgo{\epsilon^2/\kappa^2}$ (and failure probability $\bgo{\epsilon^2/\kappa^2}$) and $\ket{\by}$ is prepared with precision $\bgo{1/\epsilon^2}$, the error in the final state (compared to the ideal case) is only $\lto{1}$. Consequently, our algorithm outputs a correct $\hat{r}$ (i.e. $\abs{\hat{r}-\tau}\le \epsilon$) with probability at least $3/4-\lto{1}$.\\

\noindent\tbf{Complexity:} Since we want to estimate $r$ up to additive error $\bgo{\epsilon}$, amplitude estimation requires $\bgo{1/\epsilon}$ repetitions of the procedure $\mcal{P}$ and the procedure for preparing $\ket{\by}$. Then by Lemma \ref{lem:boostedpe}, our algorithm makes 
\bea
\bgo{ \dfrac{1}{\epsilon} \cdot \dfrac{1}{\Delta} \log{\dfrac{1}{\delta}} }
=\bgo{\dfrac{\kappa}{\epsilon} \cdot \log{\dfrac{1}{\epsilon}}}
\eea
uses of $U$. By Lemma \ref{lem:hs}, $U=e^{-i\bA}$ can be implemented with precision $\bgo{\epsilon^2/\kappa^2}$ (and failure probability $\bgo{\epsilon^2/\kappa^2}$) by a gate-efficient procedure that makes $\bgo{d^{1.5} \cdot \log{\kappa/\epsilon}}$ uses of $\px$. Meanwhile,
by Lemma \ref{lem:preparey}, $\ket{\by}$ can be prepared with precision $\bgo{\epsilon^2}$ by a gate-efficient procedure that makes $\bgo{\log{1/\epsilon}}$ uses of $\py$. As a result, this algorithm makes
\bea
\bgo{\dfrac{d^{1.5} \kappa  }{\epsilon} \cdot \poly{\log{\dfrac{\kappa}{\epsilon}}}}
\eea
uses of $\px$ and $\py$, and is gate-efficient, as claimed.
\end{proof}

Comparing Theorem \ref{thm:lrfit} and Theorem \ref{thm:lrtest}, one can see that it is easier to estimate the quality of the least-squares fit $\by \approx \bX \hat{\beta}$ than to find its parameters $\hat{\beta}=\bX^+ \by$ explicitly. Thus, in practice, we can first run the algorithm in Theorem \ref{thm:lrtest} to check whether a given data set is well-behaved (e.g. $\tau \ge 2/3$). If so, then we run the algorithm in Theorem \ref{thm:lrfit} to fit a linear regression model to this data set. The total cost of this process is dominated by that of the second stage.

\section{Lower Bound on the Complexity of Linear Regression}
\label{sec:lblr}
Our quantum algorithm for computing $\hat{\beta}=\bX^+\by$ has polynomial dependence on the condition number $\kappa$ of the design matrix $\bX$. In this section, we show that this dependence is indeed necessary. To prove this, we need the following lower bound on the quantum query complexity of a weaker version of \emph{unstructured search}.
\begin{lemma}
Let $f: \intset{N} \to \zo$ be a function such that $f(x)=1$ if and only if $x=z$ for some unknown $z \in \intset{N}$. Let $\mcal{P}_f$ be a procedure that on input $x \in \intset{N}$, outputs the value of $f(x)$. Then one has to make $\omg{\sqrt{N}/\log{N}}$ queries to $\mcal{P}_f$ to determine whether the unknown $z$ is larger than $\floor{N/2}$ or not (succeeding with probability at least $2/3$).
\label{lem:searchlb}
\end{lemma}
\begin{proof}
Suppose we can solve the given problem by making $Q$ queries to $\mcal{P}_f$. Then we can find the unknown $z$ by making $\bgo{Q \log{N}}$ queries to $\mcal{P}_f$. The idea is to use binary search. Namely, we first test whether $z$ is in the range $[0, \floor{N/2}] $ or $[\floor{N/2}+1, N]$. If the first case holds, then we test whether $z$ is in the range $[0, \floor{N/4}] $ or $[\floor{N/4}+1, \floor{N/2}]$; otherwise, we test whether $z$ is in the range $[\floor{N/2}+1, \floor{3N/4}] $ or $[\floor{3N/4}+1, N]$, and so on. We only need $\bgo{\log{N}}$ such tests to locate $z$, since each test reduces the size of candidate set by a factor of $2$. Furthermore, by assumption, each test can be accomplished by making at most $Q$ queries to $\mcal{P}_f$. Thus, we can find $z$ by making $\bgo{Q \log{N}}$ queries to $\mcal{P}_f$. On the other hand, it is known that unstructured search has quantum query complexity $\omg{\sqrt{N}}$ \cite{bennett1997strengths,boyer1998tight}. Combining these two facts, we know that $Q=\omg{\sqrt{N} / \log{N}}$. 
\end{proof}

\begin{theorem}
The LR-P problem has quantum query complexity $\omg{\kappa/\log{\kappa}}$, where $\kappa$ is the condition number of the design matrix $\bX$. 
\label{thm:lrlb}
\end{theorem}

\begin{proof}
We prove this theorem by showing that for any positive integer $N$, there exists a balanced matrix $\bX \in \bbR^{N \times 2}$ with singular values $s_1(\bX)=\tht{1/\sqrt{N}}$ and $s_2(\bX)=\tht{1}$ such that, for $\by=\frac{1}{\sqrt{N}}(1, 1, \dots, 1)^T \in \bbR^N$, $\hat{\beta}=\bX^+ \by$ is either $(1, 0)^T$ or $(0, 1)^T$, but one has to make $\omg{{\sqrt{N}}/{\log{N}}}$ queries to $\bX$ to determine which case holds (succeeding with probability at least $2/3$).

Let $\bX$ be an $N \times 2$ matrix such that its entries are all $1/\sqrt{N}$ except one entry $0$ (whose location is unknown and arbitrary). Then we know that one column of $\bX$ is equal to $\by=\frac{1}{\sqrt{N}}(1, 1, \dots, 1)^T$, and the other column of $\bX$ is linearly independent from $\by$. Consequently, using the definition
\bea
\hat{\beta}=\argmin\limits_{\beta \in \bbR^2} \nm{\bX \beta - \by},
\eea
we obtain that $\hat{\beta}$ is either $(0, 1)^T$ or $(1, 0)^T$, depending on whether the entry $0$ is in the first or second column of $\bX$, respectively. By Lemma \ref{lem:searchlb}, one must make $\omg{\sqrt{N}/\log{N}}$ queries to $\bX$ to determine which column contains the entry $0$. This implies that one also needs to make $\omg{\sqrt{N}/\log{N}}$ queries to $\bX$ to determine whether $\hat{\beta}=(0, 1)^T$ or $\hat{\beta}=(1, 0)^T$.

It remains to show that $\bX$ also satisfies the other desired properties. First, by a direct calculation, we get that $\frobnm{\bX}=\tht{1}$, $\nm{\bX}_{2,\infty}=\tht{1/\sqrt{N}}$ and hence $\sigma(\bX)=\tht{1}$. Second, note that either
\bea
\bX^T \bX = \bpm
1-\dfrac{1}{N} & 1-\dfrac{1}{N}\\
1-\dfrac{1}{N} & 1 
\epm
\eea
or
\bea
\bX^T \bX = \bpm
1 & 1-\dfrac{1}{N}\\
1-\dfrac{1}{N} & 1-\dfrac{1}{N}  
\epm.
\eea
By a direct calculation, we find that $\lambda_1(\bX^T \bX)=\tht{1/N}$
and $\lambda_2(\bX^T \bX)=\tht{1}$. It follows that $s_1(\bX)=\sqrt{\lambda_1(\bX^T \bX)}=\tht{1/\sqrt{N}}$ and $s_2(\bX)=\sqrt{\lambda_2(\bX^T \bX)}=\tht{1}$, and hence $\kappa(\bX)=\tht{\sqrt{N}}$. This concludes the proof.
\end{proof}

Clearly, the LR-P problem has time complexity $\omg{d}$, because simply writing down a $d$-dimensional vector $\beta \approx \hat{\beta}$ requires this amount of time. Combining this fact and Theorem \ref{thm:lrlb}, we know that the algorithm in Theorem \ref{thm:lrfit} cannot be dramatically improved. 

It is worth noting that Harrow, Hassidim and Lloyd (HHL) \cite{harrow2009quantum} have also given a lower bound on the quantum complexity of matrix inversion. They proved that unless $\BQP=\PSPACE$, one cannot solve the matrix inversion problem in quantum time $\kappa^{1-\delta}\cdot \poly{\log{N}}$ for some constant $\delta>0$, where $\kappa$ and $N$ are the condition number and dimension of the matrix to be inverted, respectively. We remark that this result and Theorem \ref{thm:lrlb} are incomparable. At first glance, it may seem that Theorem \ref{thm:lrlb} is stronger, since it has better dependence on $\kappa$ and it does not rely on any complexity-theoretic assumption. But recall that in our LR-P problem, we allow the design matrix X to be nonspare, while HHL only allowed sparse matrices in their work. So we only obtain a stronger bound under a stronger assumption. Nevertheless, it may be possible to use our approach to improve HHL's bound, showing that our bound holds in the sparse case as well. This is left as an interesting open question.  

\section{Discussion}
\label{sec:discuss}

To summarize, we have presented an efficient quantum algorithm for fitting a linear regression model to a given data set using the least squares approach. Different from previous algorithms which produce a quantum state encoding the optimal parameters, our algorithm outputs these numbers in the classical form. So by running it once, one completely determines the fitted model and then can use it to make predictions on new data at little cost. The running time of this algorithm is polynomial in $\log{N}$, $d$, $\kappa$ and $1/\epsilon$, where $N$ is the size of the data set, $d$ is the number of adjustable parameters, $\kappa$ is the condition number of the design matrix, and $\epsilon$ is the desired precision in the output. We also show that the polynomial dependence on $d$ and $\kappa$ is necessary. Therefore, our algorithm cannot be greatly improved. Furthermore, we also give an efficient quantum algorithm that estimates the quality of the least-squares fit (without computing its parameters explicitly). This algorithm runs faster than the one for finding this fit, and can be used to check whether the given data set qualifies for linear regression in the first place. 

One may have noticed that our algorithms actually solve two fundamental problems in linear algebra. One is to apply the pseudoinverse of a dense rectangular matrix to a vector, and the other is to estimate the norm of the projection of this vector onto the range of this matrix. Such problems frequently arise in many scenarios. So it is conceivable that our algorithms may find applications beyond linear regression.

Our algorithms might be improved in a few ways. Ambainis \cite{ambainis2012variable} proposed a technique called \emph{variable-time amplitude amplification} and utilized it to enhance the $\kappa$-dependence of HHL's algorithm \cite{harrow2009quantum} for preparing a state encoding the solution of a linear system (this techique is also used in CKS's algorithm \cite{childs2015quantum}). But it is unknown whether this technique leads to a more efficient algorithm for estimating an entry (or the difference between two entries) of this solution. If so, we would obtain a faster algorithm for fitting a linear regression model to a data set using the least squares approach. On the other hand, for estimating the quality of the fitted model, we still do not know whether the polynomial dependence on $\kappa$ is necessary. We believe that this is the case, but could not prove it. This is left as an interesting open question. 

In this paper, we have focused on linear regression with \emph{ordinary least squares} optimization (which assumes that the errors for different observations are independent). It is also worth investigating the quantum complexity of linear regression with \emph{generalized least squares} optimization (which allows the errors for different observations to be correlated). Furthermore, one might study how these complexities change when \emph{regularization} is used. For example, how hard is it to solve \emph{ridge regression} \cite{hoerl1970ridge} or \emph{Lasso} \cite{tibshirani1996regression} on a quantum computer? Finally, it would be worth exploring the power and limitation of quantum algorithms for \emph{nonlinear regression}. 

Our work is also a new contribution to the nascent field of \emph{quantum machine learning}, which has made a lot of progress in the past years  \cite{wiebe2012quantum,sentis2012quantum,briegel2012projective,paparo2012google,lloyd2013quantum,aimeur2013quantum,aaronson2015read,adcock2015advances,
ogorman2015bayesian,paparo2014quantum,rebentrost2014quantum,wiebe2014quantum,schuld2015introduction,
schuld2015simulating,sentis2015quantum,zhao2015quantum,adachi2015application,schuld2016prediction,wiebe2015quantum,biamonte2016quantum,
kerenidis2016quantum,wiebe2016quantum,dunjko2016quantum,crawford2016reinforcement,
amin2016quantum,benedetti2016quantum,benedetti2016estimation}. 
Here we briefly review this broad area, and position our work with the other works in this area (for an excellent review on quantum machine learning, see Ref.~\cite{biamonte2016quantum}). In fact, depending on the types of the learning device and the object to be learned, quantum machine learing can be divided into three branches. The first branch, which is also known as \emph{quantum-enhanced machine learning}, uses quantum mechanics to improve the performance of classical machine learning methods (e.g. \cite{wiebe2012quantum,rebentrost2014quantum,paparo2014quantum,wiebe2016quantum,schuld2016prediction}). Conversely, the second branch applies classial machine learning methods to the study of quantum systems (e.g. \cite{sentis2012quantum,wiebe2014quantum}). Finally, the third branch uses quantum approaches to study quantum systems (e.g. \cite{sentis2015quantum}). Clearly, our work is an instance of the first kind, i.e. quantum-enhanced machine learning.

Now let us look at quantum-enhanced machine learning more carefully. Traditional machine learning algorithms can be divided into three main groups based on their purpose: \emph{supervised learning} (in which an algorithm learns from example data and associated target responses that can consist of numeric values or string labels), \emph{unsupervised learning} (in which an algorithm learns from plain examples without any associated response), and \emph{reinforcement learning} (in which an agent interacts with an environment and occasionally receives rewards for its actions, which allows the agent to adapt its behavior). There has beening exciting progress in all of these three paradigms. See Refs.~\cite{wiebe2012quantum,schuld2016prediction,rebentrost2014quantum,zhao2015quantum}, Refs.~\cite{aimeur2013quantum,lloyd2013quantum,wiebe2016quantum} and Refs.~\cite{briegel2012projective,paparo2014quantum,dunjko2016quantum,crawford2016reinforcement} for examples of the first, second and third kind, respectively. Our work belongs to the first category, as it concerns least-square linear regression -- a typical supervised learning task.

Meanwhile, we can also classify the works on quantum-enhanced machine learning based on the techniques they use. It seems that most of these works fall into three groups according to this criterion. The first group use linear algebra methods (e.g. singular value decomposition), and are usually related to HHL's quantum algorithm for linear systems of equations somehow. Examples include Refs.~\cite{wiebe2012quantum,schuld2016prediction} and this work on least-squares linear regression, Ref.~\cite{rebentrost2014quantum} on support vector machine, and Ref.~\cite{zhao2015quantum} on Guassian processes. This approach could achieve exponential speedup (in some sense) over classical methods. The second group are based on amplitude amplification (including Grover's search and quantum walk). Examples include Ref.~\cite{aimeur2013quantum} on $k$-medians, Ref.~\cite{wiebe2015quantum} on $k$-nearest neighbors, Ref.~\cite{paparo2012google} on Google's PageRank, and Ref.~\cite{paparo2014quantum} on reinforcement learning. This approach usually achieves polynomial speedup over classial methods. Finally, the third group are based on quantum sampling techniques (e.g. quantum annealing). Examples include Refs.~\cite{adachi2015application,amin2016quantum,benedetti2016quantum,benedetti2016estimation,wiebe2016quantum} on (deep) Boltzmann machines. We believe that the field of quantum(-enhanced) machine learning could benefit the most from the marriage of these different ideas, and look forward to seeing more novel quantum algorithms for solving machine learning tasks. 

\section*{Acknowledgments}

The author thanks Scott Aaronson, Andrew Childs and Umesh Vazirani for helpful discussions and comments. The author also thanks the anonymous referee for providing many useful comments on an earlier version of this paper. Part of this work was done while the author was a graduate student at Computer Science Division, University of California, Berkeley. This research was supported by ARO Grant W911NF- 09-1-0440.


\begin{thebibliography}{57}%
\makeatletter
\providecommand \@ifxundefined [1]{%
 \@ifx{#1\undefined}
}%
\providecommand \@ifnum [1]{%
 \ifnum #1\expandafter \@firstoftwo
 \else \expandafter \@secondoftwo
 \fi
}%
\providecommand \@ifx [1]{%
 \ifx #1\expandafter \@firstoftwo
 \else \expandafter \@secondoftwo
 \fi
}%
\providecommand \natexlab [1]{#1}%
\providecommand \enquote  [1]{``#1''}%
\providecommand \bibnamefont  [1]{#1}%
\providecommand \bibfnamefont [1]{#1}%
\providecommand \citenamefont [1]{#1}%
\providecommand \href@noop [0]{\@secondoftwo}%
\providecommand \href [0]{\begingroup \@sanitize@url \@href}%
\providecommand \@href[1]{\@@startlink{#1}\@@href}%
\providecommand \@@href[1]{\endgroup#1\@@endlink}%
\providecommand \@sanitize@url [0]{\catcode `\\12\catcode `\$12\catcode
  `\&12\catcode `\#12\catcode `\^12\catcode `\_12\catcode `\%12\relax}%
\providecommand \@@startlink[1]{}%
\providecommand \@@endlink[0]{}%
\providecommand \url  [0]{\begingroup\@sanitize@url \@url }%
\providecommand \@url [1]{\endgroup\@href {#1}{\urlprefix }}%
\providecommand \urlprefix  [0]{URL }%
\providecommand \Eprint [0]{\href }%
\providecommand \doibase [0]{http://dx.doi.org/}%
\providecommand \selectlanguage [0]{\@gobble}%
\providecommand \bibinfo  [0]{\@secondoftwo}%
\providecommand \bibfield  [0]{\@secondoftwo}%
\providecommand \translation [1]{[#1]}%
\providecommand \BibitemOpen [0]{}%
\providecommand \bibitemStop [0]{}%
\providecommand \bibitemNoStop [0]{.\EOS\space}%
\providecommand \EOS [0]{\spacefactor3000\relax}%
\providecommand \BibitemShut  [1]{\csname bibitem#1\endcsname}%
\let\auto@bib@innerbib\@empty
\bibitem [{\citenamefont {Wiebe}\ \emph {et~al.}(2012)\citenamefont {Wiebe},
  \citenamefont {Braun},\ and\ \citenamefont {Lloyd}}]{wiebe2012quantum}%
  \BibitemOpen
  \bibfield  {author} {\bibinfo {author} {\bibfnamefont {N.}~\bibnamefont
  {Wiebe}}, \bibinfo {author} {\bibfnamefont {D.}~\bibnamefont {Braun}}, \ and\
  \bibinfo {author} {\bibfnamefont {S.}~\bibnamefont {Lloyd}},\ }\href
  {\doibase 10.1103/PhysRevLett.109.050505} {\bibfield  {journal} {\bibinfo
  {journal} {Phys. Rev. Lett.}\ }\textbf {\bibinfo {volume} {109}},\ \bibinfo
  {pages} {050505} (\bibinfo {year} {2012})}\BibitemShut {NoStop}%
\bibitem [{\citenamefont {Harrow}\ \emph {et~al.}(2009)\citenamefont {Harrow},
  \citenamefont {Hassidim},\ and\ \citenamefont {Lloyd}}]{harrow2009quantum}%
  \BibitemOpen
  \bibfield  {author} {\bibinfo {author} {\bibfnamefont {A.~W.}\ \bibnamefont
  {Harrow}}, \bibinfo {author} {\bibfnamefont {A.}~\bibnamefont {Hassidim}}, \
  and\ \bibinfo {author} {\bibfnamefont {S.}~\bibnamefont {Lloyd}},\ }\href
  {\doibase 10.1103/PhysRevLett.103.150502} {\bibfield  {journal} {\bibinfo
  {journal} {Phys. Rev. Lett.}\ }\textbf {\bibinfo {volume} {103}},\ \bibinfo
  {pages} {150502} (\bibinfo {year} {2009})}\BibitemShut {NoStop}%
\bibitem [{\citenamefont {Schuld}\ \emph {et~al.}(2016)\citenamefont {Schuld},
  \citenamefont {Sinayskiy},\ and\ \citenamefont
  {Petruccione}}]{schuld2016prediction}%
  \BibitemOpen
  \bibfield  {author} {\bibinfo {author} {\bibfnamefont {M.}~\bibnamefont
  {Schuld}}, \bibinfo {author} {\bibfnamefont {I.}~\bibnamefont {Sinayskiy}}, \
  and\ \bibinfo {author} {\bibfnamefont {F.}~\bibnamefont {Petruccione}},\
  }\href {\doibase 10.1103/PhysRevA.94.022342} {\bibfield  {journal} {\bibinfo
  {journal} {Phys. Rev. A}\ }\textbf {\bibinfo {volume} {94}},\ \bibinfo
  {pages} {022342} (\bibinfo {year} {2016})}\BibitemShut {NoStop}%
\bibitem [{\citenamefont {Lloyd}\ \emph {et~al.}(2014)\citenamefont {Lloyd},
  \citenamefont {Mohseni},\ and\ \citenamefont
  {Rebentrost}}]{lloyd2014quantum}%
  \BibitemOpen
  \bibfield  {author} {\bibinfo {author} {\bibfnamefont {S.}~\bibnamefont
  {Lloyd}}, \bibinfo {author} {\bibfnamefont {M.}~\bibnamefont {Mohseni}}, \
  and\ \bibinfo {author} {\bibfnamefont {P.}~\bibnamefont {Rebentrost}},\
  }\href@noop {} {\bibfield  {journal} {\bibinfo  {journal} {Nature Physics}\
  }\textbf {\bibinfo {volume} {10}},\ \bibinfo {pages} {631} (\bibinfo {year}
  {2014})}\BibitemShut {NoStop}%
\bibitem [{\citenamefont {Low}\ and\ \citenamefont
  {Chuang}(2016)}]{low2016hamiltonian}%
  \BibitemOpen
  \bibfield  {author} {\bibinfo {author} {\bibfnamefont {G.~H.}\ \bibnamefont
  {Low}}\ and\ \bibinfo {author} {\bibfnamefont {I.~L.}\ \bibnamefont
  {Chuang}},\ }\href@noop {} {\bibfield  {journal} {\bibinfo  {journal} {arXiv
  preprint arXiv:1610.06546}\ } (\bibinfo {year} {2016})}\BibitemShut {NoStop}%
\bibitem [{\citenamefont {Low}\ and\ \citenamefont
  {Chuang}(2017)}]{low2017optimal}%
  \BibitemOpen
  \bibfield  {author} {\bibinfo {author} {\bibfnamefont {G.~H.}\ \bibnamefont
  {Low}}\ and\ \bibinfo {author} {\bibfnamefont {I.~L.}\ \bibnamefont
  {Chuang}},\ }\href {\doibase 10.1103/PhysRevLett.118.010501} {\bibfield
  {journal} {\bibinfo  {journal} {Phys. Rev. Lett.}\ }\textbf {\bibinfo
  {volume} {118}},\ \bibinfo {pages} {010501} (\bibinfo {year}
  {2017})}\BibitemShut {NoStop}%
\bibitem [{\citenamefont {Childs}\ \emph {et~al.}(2015)\citenamefont {Childs},
  \citenamefont {Kothari},\ and\ \citenamefont {Somma}}]{childs2015quantum}%
  \BibitemOpen
  \bibfield  {author} {\bibinfo {author} {\bibfnamefont {A.~M.}\ \bibnamefont
  {Childs}}, \bibinfo {author} {\bibfnamefont {R.}~\bibnamefont {Kothari}}, \
  and\ \bibinfo {author} {\bibfnamefont {R.~D.}\ \bibnamefont {Somma}},\
  }\href@noop {} {\bibfield  {journal} {\bibinfo  {journal} {arXiv preprint
  arXiv:1511.02306}\ } (\bibinfo {year} {2015})}\BibitemShut {NoStop}%
\bibitem [{\citenamefont {Brassard}\ \emph {et~al.}(2002)\citenamefont
  {Brassard}, \citenamefont {Hoyer}, \citenamefont {Mosca},\ and\ \citenamefont
  {Tapp}}]{brassard2002quantum}%
  \BibitemOpen
  \bibfield  {author} {\bibinfo {author} {\bibfnamefont {G.}~\bibnamefont
  {Brassard}}, \bibinfo {author} {\bibfnamefont {P.}~\bibnamefont {Hoyer}},
  \bibinfo {author} {\bibfnamefont {M.}~\bibnamefont {Mosca}}, \ and\ \bibinfo
  {author} {\bibfnamefont {A.}~\bibnamefont {Tapp}},\ }\href@noop {} {\bibfield
   {journal} {\bibinfo  {journal} {Contemporary Mathematics}\ }\textbf
  {\bibinfo {volume} {305}},\ \bibinfo {pages} {53} (\bibinfo {year}
  {2002})}\BibitemShut {NoStop}%
\bibitem [{\citenamefont {Golub}\ and\ \citenamefont
  {Reinsch}(1970)}]{golub1970singular}%
  \BibitemOpen
  \bibfield  {author} {\bibinfo {author} {\bibfnamefont {G.~H.}\ \bibnamefont
  {Golub}}\ and\ \bibinfo {author} {\bibfnamefont {C.}~\bibnamefont
  {Reinsch}},\ }\href@noop {} {\bibfield  {journal} {\bibinfo  {journal}
  {Numerische mathematik}\ }\textbf {\bibinfo {volume} {14}},\ \bibinfo {pages}
  {403} (\bibinfo {year} {1970})}\BibitemShut {NoStop}%
\bibitem [{\citenamefont {Lloyd}(1996)}]{lloyd1996universal}%
  \BibitemOpen
  \bibfield  {author} {\bibinfo {author} {\bibfnamefont {S.}~\bibnamefont
  {Lloyd}},\ }\href {\doibase 10.1126/science.273.5278.1073} {\bibfield
  {journal} {\bibinfo  {journal} {Science}\ }\textbf {\bibinfo {volume}
  {273}},\ \bibinfo {pages} {1073} (\bibinfo {year} {1996})}\BibitemShut
  {NoStop}%
\bibitem [{\citenamefont {Aharonov}\ and\ \citenamefont
  {Ta-Shma}(2003)}]{aharonov2003adiabatic}%
  \BibitemOpen
  \bibfield  {author} {\bibinfo {author} {\bibfnamefont {D.}~\bibnamefont
  {Aharonov}}\ and\ \bibinfo {author} {\bibfnamefont {A.}~\bibnamefont
  {Ta-Shma}},\ }in\ \href {\doibase 10.1145/780542.780546} {\emph {\bibinfo
  {booktitle} {Proceedings of the Thirty-fifth Annual ACM Symposium on Theory
  of Computing}}},\ \bibinfo {series and number} {STOC '03}\ (\bibinfo
  {publisher} {ACM},\ \bibinfo {address} {New York, NY, USA},\ \bibinfo {year}
  {2003})\ pp.\ \bibinfo {pages} {20--29}\BibitemShut {NoStop}%
\bibitem [{\citenamefont {Childs}(2004)}]{childs2004quantum}%
  \BibitemOpen
  \bibfield  {author} {\bibinfo {author} {\bibfnamefont {A.~M.}\ \bibnamefont
  {Childs}},\ }\emph {\bibinfo {title} {Quantum information processing in
  continuous time}},\ \href@noop {} {Ph.D. thesis},\ \bibinfo  {school}
  {Massachusetts Institute of Technology} (\bibinfo {year} {2004})\BibitemShut
  {NoStop}%
\bibitem [{\citenamefont {Berry}\ \emph {et~al.}(2007)\citenamefont {Berry},
  \citenamefont {Ahokas}, \citenamefont {Cleve},\ and\ \citenamefont
  {Sanders}}]{berry2007efficient}%
  \BibitemOpen
  \bibfield  {author} {\bibinfo {author} {\bibfnamefont {D.~W.}\ \bibnamefont
  {Berry}}, \bibinfo {author} {\bibfnamefont {G.}~\bibnamefont {Ahokas}},
  \bibinfo {author} {\bibfnamefont {R.}~\bibnamefont {Cleve}}, \ and\ \bibinfo
  {author} {\bibfnamefont {B.~C.}\ \bibnamefont {Sanders}},\ }\href {\doibase
  10.1007/s00220-006-0150-x} {\bibfield  {journal} {\bibinfo  {journal}
  {Communications in Mathematical Physics}\ }\textbf {\bibinfo {volume}
  {270}},\ \bibinfo {pages} {359} (\bibinfo {year} {2007})}\BibitemShut
  {NoStop}%
\bibitem [{\citenamefont {Childs}(2010)}]{childs2010on}%
  \BibitemOpen
  \bibfield  {author} {\bibinfo {author} {\bibfnamefont {A.~M.}\ \bibnamefont
  {Childs}},\ }\href {\doibase 10.1007/s00220-009-0930-1} {\bibfield  {journal}
  {\bibinfo  {journal} {Communications in Mathematical Physics}\ }\textbf
  {\bibinfo {volume} {294}},\ \bibinfo {pages} {581} (\bibinfo {year}
  {2010})}\BibitemShut {NoStop}%
\bibitem [{\citenamefont {Childs}\ and\ \citenamefont
  {Kothari}(2011)}]{childs2011simulating}%
  \BibitemOpen
  \bibfield  {author} {\bibinfo {author} {\bibfnamefont {A.~M.}\ \bibnamefont
  {Childs}}\ and\ \bibinfo {author} {\bibfnamefont {R.}~\bibnamefont
  {Kothari}},\ }\enquote {\bibinfo {title} {Simulating sparse hamiltonians with
  star decompositions},}\ in\ \href {\doibase 10.1007/978-3-642-18073-6_8}
  {\emph {\bibinfo {booktitle} {Theory of Quantum Computation, Communication,
  and Cryptography: 5th Conference, TQC 2010, Leeds, UK, April 13-15, 2010,
  Revised Selected Papers}}},\ \bibinfo {editor} {edited by\ \bibinfo {editor}
  {\bibfnamefont {W.}~\bibnamefont {van Dam}}, \bibinfo {editor} {\bibfnamefont
  {V.~M.}\ \bibnamefont {Kendon}}, \ and\ \bibinfo {editor} {\bibfnamefont
  {S.}~\bibnamefont {Severini}}}\ (\bibinfo  {publisher} {Springer Berlin
  Heidelberg},\ \bibinfo {address} {Berlin, Heidelberg},\ \bibinfo {year}
  {2011})\ pp.\ \bibinfo {pages} {94--103}\BibitemShut {NoStop}%
\bibitem [{\citenamefont {Poulin}\ \emph {et~al.}(2011)\citenamefont {Poulin},
  \citenamefont {Qarry}, \citenamefont {Somma},\ and\ \citenamefont
  {Verstraete}}]{poulin2011quantum}%
  \BibitemOpen
  \bibfield  {author} {\bibinfo {author} {\bibfnamefont {D.}~\bibnamefont
  {Poulin}}, \bibinfo {author} {\bibfnamefont {A.}~\bibnamefont {Qarry}},
  \bibinfo {author} {\bibfnamefont {R.}~\bibnamefont {Somma}}, \ and\ \bibinfo
  {author} {\bibfnamefont {F.}~\bibnamefont {Verstraete}},\ }\href {\doibase
  10.1103/PhysRevLett.106.170501} {\bibfield  {journal} {\bibinfo  {journal}
  {Phys. Rev. Lett.}\ }\textbf {\bibinfo {volume} {106}},\ \bibinfo {pages}
  {170501} (\bibinfo {year} {2011})}\BibitemShut {NoStop}%
\bibitem [{\citenamefont {Childs}\ and\ \citenamefont
  {Wiebe}(2012)}]{childs2012hamiltonian}%
  \BibitemOpen
  \bibfield  {author} {\bibinfo {author} {\bibfnamefont {A.~M.}\ \bibnamefont
  {Childs}}\ and\ \bibinfo {author} {\bibfnamefont {N.}~\bibnamefont {Wiebe}},\
  }\href {http://dl.acm.org/citation.cfm?id=2481569.2481570} {\bibfield
  {journal} {\bibinfo  {journal} {Quantum Info. Comput.}\ }\textbf {\bibinfo
  {volume} {12}},\ \bibinfo {pages} {901} (\bibinfo {year} {2012})}\BibitemShut
  {NoStop}%
\bibitem [{\citenamefont {Berry}\ and\ \citenamefont
  {Childs}(2012)}]{berry2012blackbox}%
  \BibitemOpen
  \bibfield  {author} {\bibinfo {author} {\bibfnamefont {D.~W.}\ \bibnamefont
  {Berry}}\ and\ \bibinfo {author} {\bibfnamefont {A.~M.}\ \bibnamefont
  {Childs}},\ }\href {http://dl.acm.org/citation.cfm?id=2231036.2231040}
  {\bibfield  {journal} {\bibinfo  {journal} {Quantum Info. Comput.}\ }\textbf
  {\bibinfo {volume} {12}},\ \bibinfo {pages} {29} (\bibinfo {year}
  {2012})}\BibitemShut {NoStop}%
\bibitem [{\citenamefont {Berry}\ \emph {et~al.}(2014)\citenamefont {Berry},
  \citenamefont {Childs}, \citenamefont {Cleve}, \citenamefont {Kothari},\ and\
  \citenamefont {Somma}}]{berry2014exponential}%
  \BibitemOpen
  \bibfield  {author} {\bibinfo {author} {\bibfnamefont {D.~W.}\ \bibnamefont
  {Berry}}, \bibinfo {author} {\bibfnamefont {A.~M.}\ \bibnamefont {Childs}},
  \bibinfo {author} {\bibfnamefont {R.}~\bibnamefont {Cleve}}, \bibinfo
  {author} {\bibfnamefont {R.}~\bibnamefont {Kothari}}, \ and\ \bibinfo
  {author} {\bibfnamefont {R.~D.}\ \bibnamefont {Somma}},\ }in\ \href {\doibase
  10.1145/2591796.2591854} {\emph {\bibinfo {booktitle} {Proceedings of the
  Forty-sixth Annual ACM Symposium on Theory of Computing}}},\ \bibinfo {series
  and number} {STOC '14}\ (\bibinfo  {publisher} {ACM},\ \bibinfo {address}
  {New York, NY, USA},\ \bibinfo {year} {2014})\ pp.\ \bibinfo {pages}
  {283--292}\BibitemShut {NoStop}%
\bibitem [{\citenamefont {Berry}\ \emph
  {et~al.}(2015{\natexlab{a}})\citenamefont {Berry}, \citenamefont {Childs},
  \citenamefont {Cleve}, \citenamefont {Kothari},\ and\ \citenamefont
  {Somma}}]{berry2015simulating}%
  \BibitemOpen
  \bibfield  {author} {\bibinfo {author} {\bibfnamefont {D.~W.}\ \bibnamefont
  {Berry}}, \bibinfo {author} {\bibfnamefont {A.~M.}\ \bibnamefont {Childs}},
  \bibinfo {author} {\bibfnamefont {R.}~\bibnamefont {Cleve}}, \bibinfo
  {author} {\bibfnamefont {R.}~\bibnamefont {Kothari}}, \ and\ \bibinfo
  {author} {\bibfnamefont {R.~D.}\ \bibnamefont {Somma}},\ }\href {\doibase
  10.1103/PhysRevLett.114.090502} {\bibfield  {journal} {\bibinfo  {journal}
  {Phys. Rev. Lett.}\ }\textbf {\bibinfo {volume} {114}},\ \bibinfo {pages}
  {090502} (\bibinfo {year} {2015}{\natexlab{a}})}\BibitemShut {NoStop}%
\bibitem [{\citenamefont {Berry}\ \emph
  {et~al.}(2015{\natexlab{b}})\citenamefont {Berry}, \citenamefont {Childs},\
  and\ \citenamefont {Kothari}}]{berry2015hamiltonian}%
  \BibitemOpen
  \bibfield  {author} {\bibinfo {author} {\bibfnamefont {D.~W.}\ \bibnamefont
  {Berry}}, \bibinfo {author} {\bibfnamefont {A.~M.}\ \bibnamefont {Childs}}, \
  and\ \bibinfo {author} {\bibfnamefont {R.}~\bibnamefont {Kothari}},\ }in\
  \href {\doibase 10.1109/FOCS.2015.54} {\emph {\bibinfo {booktitle} {2015 IEEE
  56th Annual Symposium on Foundations of Computer Science}}}\ (\bibinfo {year}
  {2015})\ pp.\ \bibinfo {pages} {792--809}\BibitemShut {NoStop}%
\bibitem [{\citenamefont {Shende}\ \emph {et~al.}(2006)\citenamefont {Shende},
  \citenamefont {Bullock},\ and\ \citenamefont {Markov}}]{shende2006synthesis}%
  \BibitemOpen
  \bibfield  {author} {\bibinfo {author} {\bibfnamefont {V.~V.}\ \bibnamefont
  {Shende}}, \bibinfo {author} {\bibfnamefont {S.~S.}\ \bibnamefont {Bullock}},
  \ and\ \bibinfo {author} {\bibfnamefont {I.~L.}\ \bibnamefont {Markov}},\
  }\href {\doibase 10.1109/TCAD.2005.855930} {\bibfield  {journal} {\bibinfo
  {journal} {IEEE Transactions on Computer-Aided Design of Integrated Circuits
  and Systems}\ }\textbf {\bibinfo {volume} {25}},\ \bibinfo {pages} {1000}
  (\bibinfo {year} {2006})}\BibitemShut {NoStop}%
\bibitem [{\citenamefont {Grover}(2005)}]{grover2005fixed}%
  \BibitemOpen
  \bibfield  {author} {\bibinfo {author} {\bibfnamefont {L.~K.}\ \bibnamefont
  {Grover}},\ }\href {\doibase 10.1103/PhysRevLett.95.150501} {\bibfield
  {journal} {\bibinfo  {journal} {Phys. Rev. Lett.}\ }\textbf {\bibinfo
  {volume} {95}},\ \bibinfo {pages} {150501} (\bibinfo {year}
  {2005})}\BibitemShut {NoStop}%
\bibitem [{\citenamefont {Ambainis}(2012)}]{ambainis2012variable}%
  \BibitemOpen
  \bibfield  {author} {\bibinfo {author} {\bibfnamefont {A.}~\bibnamefont
  {Ambainis}},\ }in\ \href {\doibase
  http://dx.doi.org/10.4230/LIPIcs.STACS.2012.636} {\emph {\bibinfo {booktitle}
  {29th International Symposium on Theoretical Aspects of Computer Science
  (STACS 2012)}}},\ \bibinfo {series} {Leibniz International Proceedings in
  Informatics (LIPIcs)}, Vol.~\bibinfo {volume} {14},\ \bibinfo {editor}
  {edited by\ \bibinfo {editor} {\bibfnamefont {C.}~\bibnamefont {D{\"u}rr}}\
  and\ \bibinfo {editor} {\bibfnamefont {T.}~\bibnamefont {Wilke}}}\ (\bibinfo
  {publisher} {Schloss Dagstuhl--Leibniz-Zentrum fuer Informatik},\ \bibinfo
  {address} {Dagstuhl, Germany},\ \bibinfo {year} {2012})\ pp.\ \bibinfo
  {pages} {636--647}\BibitemShut {NoStop}%
\bibitem [{\citenamefont {Kitaev}(1995)}]{kitaev1995quantum}%
  \BibitemOpen
  \bibfield  {author} {\bibinfo {author} {\bibfnamefont {A.~Y.}\ \bibnamefont
  {Kitaev}},\ }\href@noop {} {\bibfield  {journal} {\bibinfo  {journal} {arXiv
  preprint quant-ph/9511026}\ } (\bibinfo {year} {1995})}\BibitemShut {NoStop}%
\bibitem [{\citenamefont {Cleve}\ \emph {et~al.}(1998)\citenamefont {Cleve},
  \citenamefont {Ekert}, \citenamefont {Macchiavello},\ and\ \citenamefont
  {Mosca}}]{cleve1998quantum}%
  \BibitemOpen
  \bibfield  {author} {\bibinfo {author} {\bibfnamefont {R.}~\bibnamefont
  {Cleve}}, \bibinfo {author} {\bibfnamefont {A.}~\bibnamefont {Ekert}},
  \bibinfo {author} {\bibfnamefont {C.}~\bibnamefont {Macchiavello}}, \ and\
  \bibinfo {author} {\bibfnamefont {M.}~\bibnamefont {Mosca}},\ }in\ \href@noop
  {} {\emph {\bibinfo {booktitle} {Proceedings of the Royal Society of London
  A: Mathematical, Physical and Engineering Sciences}}},\ Vol.\ \bibinfo
  {volume} {454}\ (\bibinfo {organization} {The Royal Society},\ \bibinfo
  {year} {1998})\ pp.\ \bibinfo {pages} {339--354}\BibitemShut {NoStop}%
\bibitem [{\citenamefont {Nagaj}\ \emph {et~al.}(2009)\citenamefont {Nagaj},
  \citenamefont {Wocjan},\ and\ \citenamefont {Zhang}}]{nagaj2009fast}%
  \BibitemOpen
  \bibfield  {author} {\bibinfo {author} {\bibfnamefont {D.}~\bibnamefont
  {Nagaj}}, \bibinfo {author} {\bibfnamefont {P.}~\bibnamefont {Wocjan}}, \
  and\ \bibinfo {author} {\bibfnamefont {Y.}~\bibnamefont {Zhang}},\ }\href
  {http://dl.acm.org/citation.cfm?id=2012098.2012106} {\bibfield  {journal}
  {\bibinfo  {journal} {Quantum Info. Comput.}\ }\textbf {\bibinfo {volume}
  {9}},\ \bibinfo {pages} {1053} (\bibinfo {year} {2009})}\BibitemShut
  {NoStop}%
\bibitem [{\citenamefont {Wang}(2013)}]{wang2013efficient}%
  \BibitemOpen
  \bibfield  {author} {\bibinfo {author} {\bibfnamefont {G.}~\bibnamefont
  {Wang}},\ }\href@noop {} {\bibfield  {journal} {\bibinfo  {journal} {arXiv
  preprint arXiv:1311.1851}\ } (\bibinfo {year} {2013})}\BibitemShut {NoStop}%
\bibitem [{\citenamefont {Bennett}\ \emph {et~al.}(1997)\citenamefont
  {Bennett}, \citenamefont {Bernstein}, \citenamefont {Brassard},\ and\
  \citenamefont {Vazirani}}]{bennett1997strengths}%
  \BibitemOpen
  \bibfield  {author} {\bibinfo {author} {\bibfnamefont {C.~H.}\ \bibnamefont
  {Bennett}}, \bibinfo {author} {\bibfnamefont {E.}~\bibnamefont {Bernstein}},
  \bibinfo {author} {\bibfnamefont {G.}~\bibnamefont {Brassard}}, \ and\
  \bibinfo {author} {\bibfnamefont {U.}~\bibnamefont {Vazirani}},\ }\href
  {\doibase 10.1137/S0097539796300933} {\bibfield  {journal} {\bibinfo
  {journal} {SIAM Journal on Computing}\ }\textbf {\bibinfo {volume} {26}},\
  \bibinfo {pages} {1510} (\bibinfo {year} {1997})}\BibitemShut {NoStop}%
\bibitem [{\citenamefont {Boyer}\ \emph {et~al.}(1998)\citenamefont {Boyer},
  \citenamefont {Brassard}, \citenamefont {Høyer},\ and\ \citenamefont
  {Tapp}}]{boyer1998tight}%
  \BibitemOpen
  \bibfield  {author} {\bibinfo {author} {\bibfnamefont {M.}~\bibnamefont
  {Boyer}}, \bibinfo {author} {\bibfnamefont {G.}~\bibnamefont {Brassard}},
  \bibinfo {author} {\bibfnamefont {P.}~\bibnamefont {Høyer}}, \ and\ \bibinfo
  {author} {\bibfnamefont {A.}~\bibnamefont {Tapp}},\ }\href {\doibase
  10.1002/(SICI)1521-3978(199806)46:4/5<493::AID-PROP493>3.0.CO;2-P} {\bibfield
   {journal} {\bibinfo  {journal} {Fortschritte der Physik}\ }\textbf {\bibinfo
  {volume} {46}},\ \bibinfo {pages} {493} (\bibinfo {year} {1998})}\BibitemShut
  {NoStop}%
\bibitem [{\citenamefont {Hoerl}\ and\ \citenamefont
  {Kennard}(1970)}]{hoerl1970ridge}%
  \BibitemOpen
  \bibfield  {author} {\bibinfo {author} {\bibfnamefont {A.~E.}\ \bibnamefont
  {Hoerl}}\ and\ \bibinfo {author} {\bibfnamefont {R.~W.}\ \bibnamefont
  {Kennard}},\ }\href@noop {} {\bibfield  {journal} {\bibinfo  {journal}
  {Technometrics}\ }\textbf {\bibinfo {volume} {12}},\ \bibinfo {pages} {55}
  (\bibinfo {year} {1970})}\BibitemShut {NoStop}%
\bibitem [{\citenamefont {Tibshirani}(1996)}]{tibshirani1996regression}%
  \BibitemOpen
  \bibfield  {author} {\bibinfo {author} {\bibfnamefont {R.}~\bibnamefont
  {Tibshirani}},\ }\href {http://www.jstor.org/stable/2346178} {\bibfield
  {journal} {\bibinfo  {journal} {Journal of the Royal Statistical Society.
  Series B (Methodological)}\ }\textbf {\bibinfo {volume} {58}},\ \bibinfo
  {pages} {267} (\bibinfo {year} {1996})}\BibitemShut {NoStop}%
\bibitem [{\citenamefont {Sent{\'\i}s}\ \emph {et~al.}(2012)\citenamefont
  {Sent{\'\i}s}, \citenamefont {Calsamiglia}, \citenamefont {Munoz-Tapia},\
  and\ \citenamefont {Bagan}}]{sentis2012quantum}%
  \BibitemOpen
  \bibfield  {author} {\bibinfo {author} {\bibfnamefont {G.}~\bibnamefont
  {Sent{\'\i}s}}, \bibinfo {author} {\bibfnamefont {J.}~\bibnamefont
  {Calsamiglia}}, \bibinfo {author} {\bibfnamefont {R.}~\bibnamefont
  {Munoz-Tapia}}, \ and\ \bibinfo {author} {\bibfnamefont {E.}~\bibnamefont
  {Bagan}},\ }\href@noop {} {\bibfield  {journal} {\bibinfo  {journal}
  {Scientific reports}\ }\textbf {\bibinfo {volume} {2}},\ \bibinfo {pages}
  {708} (\bibinfo {year} {2012})}\BibitemShut {NoStop}%
\bibitem [{\citenamefont {Briegel}\ and\ \citenamefont {De~las
  Cuevas}(2012)}]{briegel2012projective}%
  \BibitemOpen
  \bibfield  {author} {\bibinfo {author} {\bibfnamefont {H.~J.}\ \bibnamefont
  {Briegel}}\ and\ \bibinfo {author} {\bibfnamefont {G.}~\bibnamefont {De~las
  Cuevas}},\ }\href@noop {} {\bibfield  {journal} {\bibinfo  {journal}
  {Scientific reports}\ }\textbf {\bibinfo {volume} {2}},\ \bibinfo {pages}
  {400} (\bibinfo {year} {2012})}\BibitemShut {NoStop}%
\bibitem [{\citenamefont {{Paparo}}\ and\ \citenamefont
  {{Martin-Delgado}}(2012)}]{paparo2012google}%
  \BibitemOpen
  \bibfield  {author} {\bibinfo {author} {\bibfnamefont {G.~D.}\ \bibnamefont
  {{Paparo}}}\ and\ \bibinfo {author} {\bibfnamefont {M.~A.}\ \bibnamefont
  {{Martin-Delgado}}},\ }\href {\doibase 10.1038/srep00444} {\bibfield
  {journal} {\bibinfo  {journal} {Scientific Reports}\ }\textbf {\bibinfo
  {volume} {2}},\ \bibinfo {eid} {444} (\bibinfo {year} {2012})}\BibitemShut
  {NoStop}%
\bibitem [{\citenamefont {Lloyd}\ \emph {et~al.}(2013)\citenamefont {Lloyd},
  \citenamefont {Mohseni},\ and\ \citenamefont
  {Rebentrost}}]{lloyd2013quantum}%
  \BibitemOpen
  \bibfield  {author} {\bibinfo {author} {\bibfnamefont {S.}~\bibnamefont
  {Lloyd}}, \bibinfo {author} {\bibfnamefont {M.}~\bibnamefont {Mohseni}}, \
  and\ \bibinfo {author} {\bibfnamefont {P.}~\bibnamefont {Rebentrost}},\
  }\href@noop {} {\bibfield  {journal} {\bibinfo  {journal} {arXiv preprint
  arXiv:1307.0411}\ } (\bibinfo {year} {2013})}\BibitemShut {NoStop}%
\bibitem [{\citenamefont {A{\"i}meur}\ \emph {et~al.}(2013)\citenamefont
  {A{\"i}meur}, \citenamefont {Brassard},\ and\ \citenamefont
  {Gambs}}]{aimeur2013quantum}%
  \BibitemOpen
  \bibfield  {author} {\bibinfo {author} {\bibfnamefont {E.}~\bibnamefont
  {A{\"i}meur}}, \bibinfo {author} {\bibfnamefont {G.}~\bibnamefont
  {Brassard}}, \ and\ \bibinfo {author} {\bibfnamefont {S.}~\bibnamefont
  {Gambs}},\ }\href {\doibase 10.1007/s10994-012-5316-5} {\bibfield  {journal}
  {\bibinfo  {journal} {Machine Learning}\ }\textbf {\bibinfo {volume} {90}},\
  \bibinfo {pages} {261} (\bibinfo {year} {2013})}\BibitemShut {NoStop}%
\bibitem [{\citenamefont {Aaronson}(2015)}]{aaronson2015read}%
  \BibitemOpen
  \bibfield  {author} {\bibinfo {author} {\bibfnamefont {S.}~\bibnamefont
  {Aaronson}},\ }\href@noop {} {\bibfield  {journal} {\bibinfo  {journal}
  {Nature Physics}\ }\textbf {\bibinfo {volume} {11}},\ \bibinfo {pages} {291}
  (\bibinfo {year} {2015})}\BibitemShut {NoStop}%
\bibitem [{\citenamefont {Adcock}\ \emph {et~al.}(2015)\citenamefont {Adcock},
  \citenamefont {Allen}, \citenamefont {Day}, \citenamefont {Frick},
  \citenamefont {Hinchliff}, \citenamefont {Johnson}, \citenamefont
  {Morley-Short}, \citenamefont {Pallister}, \citenamefont {Price},\ and\
  \citenamefont {Stanisic}}]{adcock2015advances}%
  \BibitemOpen
  \bibfield  {author} {\bibinfo {author} {\bibfnamefont {J.}~\bibnamefont
  {Adcock}}, \bibinfo {author} {\bibfnamefont {E.}~\bibnamefont {Allen}},
  \bibinfo {author} {\bibfnamefont {M.}~\bibnamefont {Day}}, \bibinfo {author}
  {\bibfnamefont {S.}~\bibnamefont {Frick}}, \bibinfo {author} {\bibfnamefont
  {J.}~\bibnamefont {Hinchliff}}, \bibinfo {author} {\bibfnamefont
  {M.}~\bibnamefont {Johnson}}, \bibinfo {author} {\bibfnamefont
  {S.}~\bibnamefont {Morley-Short}}, \bibinfo {author} {\bibfnamefont
  {S.}~\bibnamefont {Pallister}}, \bibinfo {author} {\bibfnamefont
  {A.}~\bibnamefont {Price}}, \ and\ \bibinfo {author} {\bibfnamefont
  {S.}~\bibnamefont {Stanisic}},\ }\href@noop {} {\bibfield  {journal}
  {\bibinfo  {journal} {arXiv preprint arXiv:1512.02900}\ } (\bibinfo {year}
  {2015})}\BibitemShut {NoStop}%
\bibitem [{\citenamefont {O'Gorman}\ \emph {et~al.}(2015)\citenamefont
  {O'Gorman}, \citenamefont {Babbush}, \citenamefont {Perdomo-Ortiz},
  \citenamefont {Aspuru-Guzik},\ and\ \citenamefont
  {Smelyanskiy}}]{ogorman2015bayesian}%
  \BibitemOpen
  \bibfield  {author} {\bibinfo {author} {\bibfnamefont {B.}~\bibnamefont
  {O'Gorman}}, \bibinfo {author} {\bibfnamefont {R.}~\bibnamefont {Babbush}},
  \bibinfo {author} {\bibfnamefont {A.}~\bibnamefont {Perdomo-Ortiz}}, \bibinfo
  {author} {\bibfnamefont {A.}~\bibnamefont {Aspuru-Guzik}}, \ and\ \bibinfo
  {author} {\bibfnamefont {V.}~\bibnamefont {Smelyanskiy}},\ }\href@noop {}
  {\bibfield  {journal} {\bibinfo  {journal} {The European Physical Journal
  Special Topics}\ }\textbf {\bibinfo {volume} {224}},\ \bibinfo {pages} {163}
  (\bibinfo {year} {2015})}\BibitemShut {NoStop}%
\bibitem [{\citenamefont {Paparo}\ \emph {et~al.}(2014)\citenamefont {Paparo},
  \citenamefont {Dunjko}, \citenamefont {Makmal}, \citenamefont
  {Martin-Delgado},\ and\ \citenamefont {Briegel}}]{paparo2014quantum}%
  \BibitemOpen
  \bibfield  {author} {\bibinfo {author} {\bibfnamefont {G.~D.}\ \bibnamefont
  {Paparo}}, \bibinfo {author} {\bibfnamefont {V.}~\bibnamefont {Dunjko}},
  \bibinfo {author} {\bibfnamefont {A.}~\bibnamefont {Makmal}}, \bibinfo
  {author} {\bibfnamefont {M.~A.}\ \bibnamefont {Martin-Delgado}}, \ and\
  \bibinfo {author} {\bibfnamefont {H.~J.}\ \bibnamefont {Briegel}},\ }\href
  {\doibase 10.1103/PhysRevX.4.031002} {\bibfield  {journal} {\bibinfo
  {journal} {Phys. Rev. X}\ }\textbf {\bibinfo {volume} {4}},\ \bibinfo {pages}
  {031002} (\bibinfo {year} {2014})}\BibitemShut {NoStop}%
\bibitem [{\citenamefont {Rebentrost}\ \emph {et~al.}(2014)\citenamefont
  {Rebentrost}, \citenamefont {Mohseni},\ and\ \citenamefont
  {Lloyd}}]{rebentrost2014quantum}%
  \BibitemOpen
  \bibfield  {author} {\bibinfo {author} {\bibfnamefont {P.}~\bibnamefont
  {Rebentrost}}, \bibinfo {author} {\bibfnamefont {M.}~\bibnamefont {Mohseni}},
  \ and\ \bibinfo {author} {\bibfnamefont {S.}~\bibnamefont {Lloyd}},\ }\href
  {\doibase 10.1103/PhysRevLett.113.130503} {\bibfield  {journal} {\bibinfo
  {journal} {Phys. Rev. Lett.}\ }\textbf {\bibinfo {volume} {113}},\ \bibinfo
  {pages} {130503} (\bibinfo {year} {2014})}\BibitemShut {NoStop}%
\bibitem [{\citenamefont {Wiebe}\ \emph {et~al.}(2014)\citenamefont {Wiebe},
  \citenamefont {Granade}, \citenamefont {Ferrie},\ and\ \citenamefont
  {Cory}}]{wiebe2014quantum}%
  \BibitemOpen
  \bibfield  {author} {\bibinfo {author} {\bibfnamefont {N.}~\bibnamefont
  {Wiebe}}, \bibinfo {author} {\bibfnamefont {C.}~\bibnamefont {Granade}},
  \bibinfo {author} {\bibfnamefont {C.}~\bibnamefont {Ferrie}}, \ and\ \bibinfo
  {author} {\bibfnamefont {D.}~\bibnamefont {Cory}},\ }\href {\doibase
  10.1103/PhysRevA.89.042314} {\bibfield  {journal} {\bibinfo  {journal} {Phys.
  Rev. A}\ }\textbf {\bibinfo {volume} {89}},\ \bibinfo {pages} {042314}
  (\bibinfo {year} {2014})}\BibitemShut {NoStop}%
\bibitem [{\citenamefont {Schuld}\ \emph
  {et~al.}(2015{\natexlab{a}})\citenamefont {Schuld}, \citenamefont
  {Sinayskiy},\ and\ \citenamefont {Petruccione}}]{schuld2015introduction}%
  \BibitemOpen
  \bibfield  {author} {\bibinfo {author} {\bibfnamefont {M.}~\bibnamefont
  {Schuld}}, \bibinfo {author} {\bibfnamefont {I.}~\bibnamefont {Sinayskiy}}, \
  and\ \bibinfo {author} {\bibfnamefont {F.}~\bibnamefont {Petruccione}},\
  }\href {\doibase 10.1080/00107514.2014.964942} {\bibfield  {journal}
  {\bibinfo  {journal} {Contemporary Physics}\ }\textbf {\bibinfo {volume}
  {56}},\ \bibinfo {pages} {172} (\bibinfo {year}
  {2015}{\natexlab{a}})}\BibitemShut {NoStop}%
\bibitem [{\citenamefont {Schuld}\ \emph
  {et~al.}(2015{\natexlab{b}})\citenamefont {Schuld}, \citenamefont
  {Sinayskiy},\ and\ \citenamefont {Petruccione}}]{schuld2015simulating}%
  \BibitemOpen
  \bibfield  {author} {\bibinfo {author} {\bibfnamefont {M.}~\bibnamefont
  {Schuld}}, \bibinfo {author} {\bibfnamefont {I.}~\bibnamefont {Sinayskiy}}, \
  and\ \bibinfo {author} {\bibfnamefont {F.}~\bibnamefont {Petruccione}},\
  }\href {\doibase http://dx.doi.org/10.1016/j.physleta.2014.11.061} {\bibfield
   {journal} {\bibinfo  {journal} {Physics Letters A}\ }\textbf {\bibinfo
  {volume} {379}},\ \bibinfo {pages} {660 } (\bibinfo {year}
  {2015}{\natexlab{b}})}\BibitemShut {NoStop}%
\bibitem [{\citenamefont {Sent{\'\i}s}\ \emph {et~al.}(2015)\citenamefont
  {Sent{\'\i}s}, \citenamefont {Gu{\c{t}}{\u{a}}},\ and\ \citenamefont
  {Adesso}}]{sentis2015quantum}%
  \BibitemOpen
  \bibfield  {author} {\bibinfo {author} {\bibfnamefont {G.}~\bibnamefont
  {Sent{\'\i}s}}, \bibinfo {author} {\bibfnamefont {M.}~\bibnamefont
  {Gu{\c{t}}{\u{a}}}}, \ and\ \bibinfo {author} {\bibfnamefont
  {G.}~\bibnamefont {Adesso}},\ }\href@noop {} {\bibfield  {journal} {\bibinfo
  {journal} {EPJ Quantum Technology}\ }\textbf {\bibinfo {volume} {2}},\
  \bibinfo {pages} {17} (\bibinfo {year} {2015})}\BibitemShut {NoStop}%
\bibitem [{\citenamefont {Zhao}\ \emph {et~al.}(2015)\citenamefont {Zhao},
  \citenamefont {Fitzsimons},\ and\ \citenamefont
  {Fitzsimons}}]{zhao2015quantum}%
  \BibitemOpen
  \bibfield  {author} {\bibinfo {author} {\bibfnamefont {Z.}~\bibnamefont
  {Zhao}}, \bibinfo {author} {\bibfnamefont {J.~K.}\ \bibnamefont
  {Fitzsimons}}, \ and\ \bibinfo {author} {\bibfnamefont {J.~F.}\ \bibnamefont
  {Fitzsimons}},\ }\href@noop {} {\bibfield  {journal} {\bibinfo  {journal}
  {arXiv preprint arXiv:1512.03929}\ } (\bibinfo {year} {2015})}\BibitemShut
  {NoStop}%
\bibitem [{\citenamefont {Adachi}\ and\ \citenamefont
  {Henderson}(2015)}]{adachi2015application}%
  \BibitemOpen
  \bibfield  {author} {\bibinfo {author} {\bibfnamefont {S.~H.}\ \bibnamefont
  {Adachi}}\ and\ \bibinfo {author} {\bibfnamefont {M.~P.}\ \bibnamefont
  {Henderson}},\ }\href@noop {} {\bibfield  {journal} {\bibinfo  {journal}
  {arXiv preprint arXiv:1510.06356}\ } (\bibinfo {year} {2015})}\BibitemShut
  {NoStop}%
\bibitem [{\citenamefont {Wiebe}\ \emph {et~al.}(2015)\citenamefont {Wiebe},
  \citenamefont {Kapoor},\ and\ \citenamefont {Svore}}]{wiebe2015quantum}%
  \BibitemOpen
  \bibfield  {author} {\bibinfo {author} {\bibfnamefont {N.}~\bibnamefont
  {Wiebe}}, \bibinfo {author} {\bibfnamefont {A.}~\bibnamefont {Kapoor}}, \
  and\ \bibinfo {author} {\bibfnamefont {K.~M.}\ \bibnamefont {Svore}},\
  }\href@noop {} {\bibfield  {journal} {\bibinfo  {journal} {Quantum
  Information and Computation}\ }\textbf {\bibinfo {volume} {15}} (\bibinfo
  {year} {2015})}\BibitemShut {NoStop}%
\bibitem [{\citenamefont {Biamonte}\ \emph {et~al.}(2016)\citenamefont
  {Biamonte}, \citenamefont {Wittek}, \citenamefont {Pancotti}, \citenamefont
  {Rebentrost}, \citenamefont {Wiebe},\ and\ \citenamefont
  {Lloyd}}]{biamonte2016quantum}%
  \BibitemOpen
  \bibfield  {author} {\bibinfo {author} {\bibfnamefont {J.}~\bibnamefont
  {Biamonte}}, \bibinfo {author} {\bibfnamefont {P.}~\bibnamefont {Wittek}},
  \bibinfo {author} {\bibfnamefont {N.}~\bibnamefont {Pancotti}}, \bibinfo
  {author} {\bibfnamefont {P.}~\bibnamefont {Rebentrost}}, \bibinfo {author}
  {\bibfnamefont {N.}~\bibnamefont {Wiebe}}, \ and\ \bibinfo {author}
  {\bibfnamefont {S.}~\bibnamefont {Lloyd}},\ }\href@noop {} {\bibfield
  {journal} {\bibinfo  {journal} {arXiv preprint arXiv:1611.09347}\ } (\bibinfo
  {year} {2016})}\BibitemShut {NoStop}%
\bibitem [{\citenamefont {Kerenidis}\ and\ \citenamefont
  {Prakash}(2016)}]{kerenidis2016quantum}%
  \BibitemOpen
  \bibfield  {author} {\bibinfo {author} {\bibfnamefont {I.}~\bibnamefont
  {Kerenidis}}\ and\ \bibinfo {author} {\bibfnamefont {A.}~\bibnamefont
  {Prakash}},\ }\href@noop {} {\bibfield  {journal} {\bibinfo  {journal} {arXiv
  preprint arXiv:1603.08675}\ } (\bibinfo {year} {2016})}\BibitemShut {NoStop}%
\bibitem [{\citenamefont {Wiebe}\ \emph {et~al.}(2016)\citenamefont {Wiebe},
  \citenamefont {Kapoor},\ and\ \citenamefont {Svore}}]{wiebe2016quantum}%
  \BibitemOpen
  \bibfield  {author} {\bibinfo {author} {\bibfnamefont {N.}~\bibnamefont
  {Wiebe}}, \bibinfo {author} {\bibfnamefont {A.}~\bibnamefont {Kapoor}}, \
  and\ \bibinfo {author} {\bibfnamefont {K.~M.}\ \bibnamefont {Svore}},\
  }\href@noop {} {\bibfield  {journal} {\bibinfo  {journal} {Quantum
  Information and Computation}\ }\textbf {\bibinfo {volume} {16}} (\bibinfo
  {year} {2016})}\BibitemShut {NoStop}%
\bibitem [{\citenamefont {Dunjko}\ \emph {et~al.}(2016)\citenamefont {Dunjko},
  \citenamefont {Taylor},\ and\ \citenamefont {Briegel}}]{dunjko2016quantum}%
  \BibitemOpen
  \bibfield  {author} {\bibinfo {author} {\bibfnamefont {V.}~\bibnamefont
  {Dunjko}}, \bibinfo {author} {\bibfnamefont {J.~M.}\ \bibnamefont {Taylor}},
  \ and\ \bibinfo {author} {\bibfnamefont {H.~J.}\ \bibnamefont {Briegel}},\
  }\href {\doibase 10.1103/PhysRevLett.117.130501} {\bibfield  {journal}
  {\bibinfo  {journal} {Phys. Rev. Lett.}\ }\textbf {\bibinfo {volume} {117}},\
  \bibinfo {pages} {130501} (\bibinfo {year} {2016})}\BibitemShut {NoStop}%
\bibitem [{\citenamefont {Crawford}\ \emph {et~al.}(2016)\citenamefont
  {Crawford}, \citenamefont {Levit}, \citenamefont {Ghadermarzy}, \citenamefont
  {Oberoi},\ and\ \citenamefont {Ronagh}}]{crawford2016reinforcement}%
  \BibitemOpen
  \bibfield  {author} {\bibinfo {author} {\bibfnamefont {D.}~\bibnamefont
  {Crawford}}, \bibinfo {author} {\bibfnamefont {A.}~\bibnamefont {Levit}},
  \bibinfo {author} {\bibfnamefont {N.}~\bibnamefont {Ghadermarzy}}, \bibinfo
  {author} {\bibfnamefont {J.~S.}\ \bibnamefont {Oberoi}}, \ and\ \bibinfo
  {author} {\bibfnamefont {P.}~\bibnamefont {Ronagh}},\ }\href@noop {}
  {\bibfield  {journal} {\bibinfo  {journal} {arXiv preprint arXiv:1612.05695}\
  } (\bibinfo {year} {2016})}\BibitemShut {NoStop}%
\bibitem [{\citenamefont {Amin}\ \emph {et~al.}(2016)\citenamefont {Amin},
  \citenamefont {Andriyash}, \citenamefont {Rolfe}, \citenamefont
  {Kulchytskyy},\ and\ \citenamefont {Melko}}]{amin2016quantum}%
  \BibitemOpen
  \bibfield  {author} {\bibinfo {author} {\bibfnamefont {M.~H.}\ \bibnamefont
  {Amin}}, \bibinfo {author} {\bibfnamefont {E.}~\bibnamefont {Andriyash}},
  \bibinfo {author} {\bibfnamefont {J.}~\bibnamefont {Rolfe}}, \bibinfo
  {author} {\bibfnamefont {B.}~\bibnamefont {Kulchytskyy}}, \ and\ \bibinfo
  {author} {\bibfnamefont {R.}~\bibnamefont {Melko}},\ }\href@noop {}
  {\bibfield  {journal} {\bibinfo  {journal} {arXiv preprint arXiv:1601.02036}\
  } (\bibinfo {year} {2016})}\BibitemShut {NoStop}%
\bibitem [{\citenamefont {Benedetti}\ \emph
  {et~al.}(2016{\natexlab{a}})\citenamefont {Benedetti}, \citenamefont
  {Realpe-G{\'o}mez}, \citenamefont {Biswas},\ and\ \citenamefont
  {Perdomo-Ortiz}}]{benedetti2016quantum}%
  \BibitemOpen
  \bibfield  {author} {\bibinfo {author} {\bibfnamefont {M.}~\bibnamefont
  {Benedetti}}, \bibinfo {author} {\bibfnamefont {J.}~\bibnamefont
  {Realpe-G{\'o}mez}}, \bibinfo {author} {\bibfnamefont {R.}~\bibnamefont
  {Biswas}}, \ and\ \bibinfo {author} {\bibfnamefont {A.}~\bibnamefont
  {Perdomo-Ortiz}},\ }\href@noop {} {\bibfield  {journal} {\bibinfo  {journal}
  {arXiv preprint arXiv:1609.02542}\ } (\bibinfo {year}
  {2016}{\natexlab{a}})}\BibitemShut {NoStop}%
\bibitem [{\citenamefont {Benedetti}\ \emph
  {et~al.}(2016{\natexlab{b}})\citenamefont {Benedetti}, \citenamefont
  {Realpe-G\'omez}, \citenamefont {Biswas},\ and\ \citenamefont
  {Perdomo-Ortiz}}]{benedetti2016estimation}%
  \BibitemOpen
  \bibfield  {author} {\bibinfo {author} {\bibfnamefont {M.}~\bibnamefont
  {Benedetti}}, \bibinfo {author} {\bibfnamefont {J.}~\bibnamefont
  {Realpe-G\'omez}}, \bibinfo {author} {\bibfnamefont {R.}~\bibnamefont
  {Biswas}}, \ and\ \bibinfo {author} {\bibfnamefont {A.}~\bibnamefont
  {Perdomo-Ortiz}},\ }\href {\doibase 10.1103/PhysRevA.94.022308} {\bibfield
  {journal} {\bibinfo  {journal} {Phys. Rev. A}\ }\textbf {\bibinfo {volume}
  {94}},\ \bibinfo {pages} {022308} (\bibinfo {year}
  {2016}{\natexlab{b}})}\BibitemShut {NoStop}%
\end{thebibliography}

%

\end{document}